\documentclass[12pt]{article}
\usepackage{dcolumn}
\usepackage{multicol}
\usepackage{multirow}
\usepackage{array}

\usepackage{jheppub} 
\usepackage{amsmath}
\usepackage{braket}
\usepackage{float}
\usepackage{slashed}
\usepackage{graphicx} 
\usepackage{subcaption}
\usepackage{mathtools}
\usepackage{color}
\usepackage{breqn}
\usepackage{hyperref}
\usepackage{mathtools, breqn, autobreak}
\allowdisplaybreaks[1]

\numberwithin{equation}{section}
\newcommand{\be}{\begin{equation}}
\newcommand{\ee}{\end{equation}}
\newcommand{\beq}{\begin{equation}}
\newcommand{\eeq}{\end{equation}}
\newcommand{\bea}{\begin{eqnarray}}
\newcommand{\eea}{\end{eqnarray}}
\newcommand{\bear}{\begin{eqnarray}}
\newcommand{\eear}{\end{eqnarray}}

\newcommand{\ba}{\begin{array}}
\newcommand{\ea}{\end{array}}

\def\ltap{\ \raise.3ex\hbox{$<$\kern-.75em\lower1ex\hbox{$\sim$}}\ }
\def\gtap{\ \raise.3ex\hbox{$>$\kern-.75em\lower1ex\hbox{$\sim$}}\ }
\def\lsim{\ \raise.3ex\hbox{$<$\kern-.75em\lower1ex\hbox{$\sim$}}\ }
\def\gsim{\ \raise.3ex\hbox{$>$\kern-.75em\lower1ex\hbox{$\sim$}}\ }

\title{\vskip-3cm{\baselineskip14pt
\centerline{\hfill\rm\normalsize FERMILAB-PUB-26-0421-T}
\vskip0.2cm
\centerline{\hfill\rm\normalsize
\today}
}
\vskip1.5cm
\boldmath \large Neutrino Dipole Moments and Radiative Signatures from Partial Compositeness}

\author[a,b]{Beno\^it Assi}
\author[b]{and Pedro~A.N.~Machado}

\affiliation[a]{Department of Physics, University of Cincinnati, Cincinnati, Ohio 45221, USA}
\affiliation[b]{Particle Theory Department, Fermilab, Batavia, IL 60510, USA}
\emailAdd{assibt@ucmail.uc.edu}
\emailAdd{pmachado@fnal.gov}

\bigskip

\abstract{
We investigate composite neutrino models where heavy neutrinos emerge as bound states from a near-conformal strongly coupled sector.
Standard Model neutrinos mix with these composite singlets via an inverse seesaw mechanism, where the anomalous scaling dimensions of the composite-sector operators naturally suppress light neutrino masses to sub-eV scales.
Matching the conformal dynamics onto low-energy theory yields enhanced electromagnetic transition dipole operators with couplings \(d_{\mu N} \sim 10^{-6}\)--\(10^{-8}\,\mathrm{GeV}^{-1}\), parametrically larger than the loop-level predictions of minimal Dirac or Majorana models.
We carry out a dedicated event-level simulation of the production-and-decay chain \(\nu X \to \mathcal{U} X \to \nu\gamma X\) and compute the resulting event rates at MiniBooNE and MINERvA within the model, accounting for the composite production cross section and decay kinematics in detail.
We further present predictions for the photon energy, angular, and multiplicity distributions. 
For the benchmark scenarios accessible at these experiments the radiative signal is predominantly single-photon; the composite structure additionally permits fragmentation of the up-scattered state into multiple heavy neutrinos, each decaying as $N\to\nu\gamma$, with multi-photon final states emerging  for lighter compositeness scales or higher beam energies as a qualitatively new probe of the composite dynamics.}
\begin{document} 
\maketitle

\flushbottom

\section{Introduction}

Neutrino physics has entered a precision era, yet many fundamental questions remain unanswered. 
The observation of neutrino oscillations unambiguously indicates that neutrinos possess mass, a fact that the Standard Model (SM) cannot naturally accommodate without introducing \emph{ad hoc} right-handed neutrino fields. 
The na\"ive Dirac mass mechanism would require Yukawa couplings of order $10^{-12}$ or smaller, an unnatural fine-tuning that begs for a dynamical explanation. 
Moreover, since the right-handed neutrino would be a standard model singlet, an unbroken global lepton number symmetry would need to be postulated to avoid a Majorana mass term in that sector.
Speculatively, gravity effects could break this global lepton number symmetry~\cite{Coleman:1988cy, Giddings:1988cx, Giddings:1988wv, Abbott:1989jw}, leading to a Planck-mass suppressed Weinberg operator~\cite{Weinberg:1979sa}.
This would be ruled out by data, as it would contribute to neutrino masses leading to pseudo-Dirac neutrinos with relatively large mass splittings~\cite{Wolfenstein:1981kw, Petcov:1982ya, Bilenky:1983wt, deGouvea:2009fp}.
Regardless, the Higgs mechanism, which successfully generates masses for quarks and charged leptons, offers no insight into why neutrino masses should be so suppressed.

This extreme parametric smallness of the neutrino Yukawa coupling, coupled with the absence of direct evidence for new physics at the electroweak scale, has motivated a resurgence of interest in compositeness scenarios~\cite{Kaplan:1983fs, Manohar:1983md, Dugan:1984hq, Kaplan:1991dc, Contino:2006qr, Gavela:2014uta, Guo:2015isa, Buchalla:2020kdh, Brivio:2025yrr}. 
Compositeness offers an elegant framework for addressing naturalness questions: if particles that appear elementary at accessible energies are actually bound states of more fundamental constituents, hierarchies in mass scales can emerge naturally from the dynamics of the underlying strong interactions. 
Just as the pion mass is parametrically suppressed relative to the confinement scale in QCD, composite leptons could naturally explain the observed mass hierarchies without fine-tuning. 
Furthermore, compositeness provides a compelling alternative to conventional seesaw mechanisms, which typically invoke very heavy right-handed neutrinos at scales inaccessible to experiment, potentially inducing yet another hierarchy problem with the Higgs~\cite{Vissani:1997ys}, by allowing new physics to appear at lower, potentially testable energy scales.

Several models of lepton compositeness have been explored in the literature, ranging from Technicolor-inspired scenarios and composite Higgs models with partially composite leptons, to models where only neutrinos are composite while charged leptons remain elementary, see e.g. Refs.~\cite{Appelquist:2002me, Banks:2010zn}. 
Renormalizable approaches to quark and lepton compositeness provide an alternative framework where compositeness emerges at higher scales with different phenomenological implications~\cite{Dobrescu:2021fny, Assi:2022jwg, Assi:2025rjx}. 
Models with composite right-handed or sterile neutrinos have explored various mechanisms for generating small neutrino masses, including connections to dark matter and cosmology~\cite{ArkaniHamed:1998pf, Gherghetta:2003hf, vonGersdorff:2008is, Grossman:2008xb, Grossman:2010iq, McDonald:2010jm, Robinson:2012wu, Agashe:2015izu, Chakraborty:2021fkp, Ahmed:2023vdb}. 
Each approach faces distinct challenges: maintaining custodial symmetry, avoiding excessive flavor-changing neutral currents, and ensuring compatibility with precision electroweak observables. 
A promising avenue involves partial compositeness, where SM fermions mix with composite operators from a strongly-coupled sector, allowing for a controlled departure from full compositeness while preserving the key phenomenological advantages. 
Various aspects of composite dark matter and composite unification have demonstrated the viability of such scenarios in related contexts~\cite{Fox:2012ee,Ghosh:2014ida}.

In this work, we adopt the framework of low-scale partial compositeness with a near-conformal sector proposed in Ref.~\cite{Chacko:2020zze}, treating heavy neutrinos as composite states emerging from a strongly coupled CFT-like theory~\cite{Georgi:2007ek, Georgi:2007si, Grinstein:2008qk, Kenzie-thesis}. Above the compositeness scale, the spectrum of these states blends into the continuum characteristic of unparticle physics, while at and below the compositeness scale they are well-defined particles; we use ``unparticle'' only when referring to the conformal-phase production object.
This approach offers several distinctive features over alternative compositeness scenarios. 
First, the nontrivial scaling dimensions of composite operators in the 
near-conformal regime provide a dynamical mechanism for suppressing neutrino 
masses, replacing arbitrarily small Yukawa couplings with calculable anomalous 
dimensions~\cite{Altmannshofer:2014vra, Ghosh:2014ida}. 
Second, 
the composite sector can naturally reside 
at scales as low as a few~TeV, making the scenario directly testable at current 
and upcoming collider experiments. 
Third, as we will show in this paper, the composite nature of heavy neutrinos can lead to sizable dipole-moment 
operators, providing unique signatures in radiative decays and scattering 
processes that are parametrically larger than in minimal Dirac or Majorana 
models~\cite{Gelmini:2004ah}.
Finally, the model simultaneously addresses neutrino mass generation and the 
origin of their extreme parametric smallness, and may also offer implications 
for short-baseline anomalies and collider searches, providing a coherent 
alternative to disjoint explanations. 
The connection between composite sectors and neutrino phenomenology has been explored in various contexts~\cite{Ghosh:2014ida, Mondal:2015zba, Fukuda:2018oco}, and our framework extends these ideas to incorporate enhanced electromagnetic interactions. We tie the transition dipole to the same composite operator that generates the inverse-seesaw neutrino mass, retain the full unparticle spectral structure of the produced state, and carry the prediction through to event-level observables and estimated sensitivities at neutrino beam experiments.

The main finding of Ref.~\cite{Chacko:2020zze} is the realization of the inverse 
seesaw mechanism within the composite sector, see also Ref.~\cite{Agashe:2015izu}. 
Standard model neutrinos mix with composite singlet states, generating small 
Majorana masses without requiring extremely heavy scales. 
The matching of the UV conformal dynamics onto a low-energy effective field 
theory (EFT) reveals that interactions between the SM and the composite sector 
are encoded in higher-dimensional operators with nontrivial scaling.

As shown in Ref.~\cite{Chacko:2020zze}, the inverse seesaw preference points 
toward a relatively low compositeness scale. 
If this is the case, one naturally expects additional operators beyond those 
explicitly analyzed there. 
In this work we explore the possibility of a transition magnetic
moment operator coupling heavy composite neutrinos to photons,
finding that the resulting dipole interactions can be significantly enhanced,
yielding effective couplings in the range
$10^{-6}$--$10^{-8}\,\text{GeV}^{-1}$, well above those in minimal Dirac or
Majorana scenarios and within reach of current experiments such as SBND, MicroBooNE, and ICARUS~\cite{MicroBooNE:2015bmn, MicroBooNE:2016pwy, Machado:2019oxb, SBND:2025plf}.

Concretely, this work makes two main contributions. First, we show that the same composite operator responsible for the inverse-seesaw neutrino mass generates a transition dipole moment that is parametrically enhanced relative to minimal models, and we map the predicted parameter space as a function of the compositeness scale and UV cutoff. Second, we perform a dedicated event-level simulation of the coherent production and radiative decay \(\nu X \to \mathcal{U} X \to \nu\gamma X\) and use it to estimate the MiniBooNE and MINERvA sensitivities \emph{within} the composite model, together with predictions for the photon energy, angular, multiplicity, and \(E\theta^2\) distributions relevant to short-baseline searches.

This paper is organized as follows. Section~\ref{sec:inv_see} reviews the inverse seesaw mechanism with partial compositeness and establishes our notation. Section~\ref{sec:dipole} derives the transition and direct dipole moment operators from the UV theory and develops the production and decay formalism. Section~\ref{sec:pheno} describes the event-level simulation and presents the resulting parameter space in comparison with current experimental, cosmological, astrophysical, and collider constraints. Section~\ref{sec:sign} discusses the kinematic signatures and how liquid-argon time projection chambers can probe them. We conclude in Section~\ref{sec:conclusion} with a summary and outlook for future investigations.


\section{Inverse See-Saw Model with Partial Compositeness}
\label{sec:inv_see}

Following Ref.~\cite{Chacko:2020zze}, we adopt a framework in which neutrinos obtain their masses through partial compositeness. 
We briefly review the framework of Ref.~\cite{Chacko:2020zze} before laying out the new dipole operators considered here.
The strong dynamics is modelled as a near-conformal field theory that confines at a scale \(\Lambda\), generating composite fermion pairs \(N\) and \(N^c\) with masses \(M_N \sim \Lambda\). Throughout, we assume that the composite fermion \(N\) is the lightest state in the hidden sector, so that its only kinematically allowed decays are to Standard Model final states via the neutrino portal. This is essential for the radiative-decay phenomenology developed below: if instead the lightest composite state were a scalar (e.g.\ a dilaton), the dominant signatures would be those of a long-lived scalar rather than the dipole-induced \(N\to\nu\gamma\) decay, a qualitatively different scenario studied in Ref.~\cite{Ahmed:2025ldh}. The connection to the Standard Model is established through a higher-dimensional operator in the UV theory:
\begin{equation}
\mathcal{L}_{\rm UV} \supset \frac{\hat{\lambda}}{M_{\mathrm{UV}}^{\Delta_{\mathrm{N}} - 3/2}} \, L H \mathcal{O}_{\rm N} + {\rm h.c.} \;,
\label{eq:LagUVss}
\end{equation}
where \(L\) and \(H\) are the SM lepton and Higgs doublets, \(\mathcal{O}_{\rm N}\) is a composite operator with scaling dimension \(\Delta_{\mathrm{N}}\), and \(\hat{\lambda}\) is a dimensionless coupling of order unity.

Unitarity requires \(\Delta_{\mathrm{N}} \ge 3/2\) for fermionic operators.  
We further restrict to \(3/2 \le \Delta_{\mathrm{N}} < 5/2\) because the fermionic unparticle propagator admits a spectral representation
\[
\int d^4x\,e^{ipx}\langle 0|T\left[ \mathcal{O}_{\rm N}(x) \mathcal{O}_{\rm N}^\dagger(0)\right] |0 \rangle
=\int_0^\infty dM^2\,\rho(M^2)\,
\frac{i\,\sigma^\mu p_\mu}{p^2-M^2+i\epsilon},
\]
with a spectral density scaling as \(\rho(M^2)\propto (M^2)^{\Delta_{\mathrm{N}}-5/2}\).  
For \(\Delta_{\mathrm{N}}\ge 5/2\), the integral over large \(M^2\) diverges, and once \(\mathcal{O}_{\rm N}\) couples to SM fields these UV divergences require additional counterterms involving SM operators alone, since the divergent contributions arise from the high-\(M^2\) tail of the spectral integral after closing the CFT legs.
Restricting to \(\Delta_{\mathrm{N}}<5/2\) avoids these extra operators and keeps the portal EFT self-contained and calculable.

Within this range, the two-point function of \(\mathcal{O}_{\rm N}\) takes the form
\begin{equation}
\int d^4 x\, e^{ipx}\,
\langle 0|T\!\left[ \mathcal{O}_{\rm N}(x)\mathcal{O}_{\rm N}^\dagger(0)\right]|0\rangle
=
\frac{A_{\Delta_{\mathrm{N}}-1/2}}{2 i\,\cos(\Delta_{\mathrm{N}}\pi)}
\frac{\sigma^\mu p_\mu}{\big(-p^2 - i\epsilon\big)^{5/2-\Delta_{\mathrm{N}}}},
\end{equation}
where the coefficient \(A_{\Delta_{\mathrm{N}}-1/2}\) follows from Georgi’s
fractional-phase-space construction for \(n\)-body phase space with non-integer \(n\)
\cite{Georgi:2007si}:
\begin{equation}\label{eq:A_delta}
A_{\Delta_{\mathrm{N}} - 1/2} =
\frac{16\pi^{5/2}}{(2\pi)^{\,2\Delta_{\mathrm{N}}-1}}
\frac{\Gamma(\Delta_{\mathrm{N}})}
     {\Gamma(\Delta_{\mathrm{N}} - 3/2)\,
      \Gamma(2\Delta_{\mathrm{N}} - 1)}\,.
\end{equation}
This normalization determines the matching onto low-energy observables. At the confinement scale, the UV operator flows to
\begin{equation}
\mathcal{L}_{\rm IR} \supset \lambda\, L H N + {\rm h.c.} \,,
\end{equation}
where
\begin{equation}
\lambda \sim C_\lambda\, \hat{\lambda}\,\left(\frac{\Lambda}{M_{\mathrm{UV}}}\right)^{\Delta_{\mathrm{N}}-3/2}\,,
\quad
C_\lambda = \frac{(4\pi)^{3/2-\Delta_{\mathrm{N}}}}{\Gamma(\Delta_{\mathrm{N}}-3/2)}
\sqrt{\frac{ \pi}{ (\Delta_{\mathrm{N}}-3/2)\cos (\Delta_{\mathrm{N}}\pi) }} \,.
\label{eq:matching_lambda}
\end{equation}

Lepton number violation enters via a scalar operator \(\mathcal{O}_{\rm 2N^c}\) with scaling dimension \(\Delta_{\rm 2N^c}\),
\begin{equation}
\mathcal{L}_{\rm UV} \supset \frac{\hat{\mu}^c}{M_{\mathrm{UV}}^{\Delta_{\rm 2N^c} - 4}}\, \mathcal{O}_{\rm 2N^c} + {\rm h.c.}  ,
\end{equation}
where $\hat\mu^c$ is an $\mathcal{O}(1)$ dimensionless coefficient.
Unitarity requires \(\Delta_{\rm 2N^c} \geq 1\) for scalars.
We normalize the operator \(\mathcal{O}_{\rm 2N^c}\) using the absorptive (imaginary) part of its two-point function, which for the range of scaling dimensions of interest takes the form
\begin{equation}
\int d^4 x\, e^{ipx}\,
\langle 0|T\left[ \mathcal{O}_{\rm 2N^c}^{\dagger}(x) \mathcal{O}_{\rm 2N^c}(0)\right] |0 \rangle
\Big|_{\rm abs.} = \frac{1}{2}\, A_{\Delta_{\rm 2N^c}}\,
\big(p^2\big)^{\Delta_{\rm 2N^c}-2}\,\theta(p^2)~,
\end{equation}
where $A_{\Delta_{\rm 2N^c}}$ is the scalar normalization constant of Refs.~\cite{Georgi:2007ek,Georgi:2007si,Chacko:2020zze,Ahmed:2025ldh}, analogous to but distinct from the fermionic constant in Eq.~\eqref{eq:A_delta}.
The absorptive part is ultraviolet safe for the entire range \(\Delta_{\rm 2N^c} > 1\), and is unaffected by the regularization needed to define the full correlator for \(\Delta_{\rm 2N^c} \geq 2\)~\cite{Georgi:2007ek,Georgi:2007si,Chacko:2020zze,Ahmed:2025ldh}. Since only this normalization enters our results, it suffices to define the operator over the whole range \(\Delta_{\rm 2N^c} > 1\), as in Refs.~\cite{Chacko:2020zze,Ahmed:2025ldh}.~\footnote{We thank Zackaria Chacko for explaining this argument to us.}

The operator \(\mathcal{O}_{\rm 2N^c}\) generates a Majorana mass at low energies,
\begin{equation}
\mathcal{L}_{\rm IR} \supset \frac{\mu^c}{2} \left(N^c\right)^2 + {\rm h.c.},
\end{equation}
with IR coefficients,
\begin{equation}
    \mu^c \sim C_\mu \hat{\mu}^c\, \Lambda \left(\frac{\Lambda}{M_{\mathrm{UV}}} 
\right)^{\Delta_{\rm 2N^c} - 4},
\quad
C_\mu = \frac{(4\pi)^{2-\Delta_{2\mathrm{N}^c}}}{\Gamma(\Delta_{2\mathrm{N}^c}-1)}
\sqrt{\frac{ 1}{ \Delta_{2\mathrm{N}^c}-1 }} \,.
\end{equation}
Integrating out the heavy states yields light neutrino masses via the inverse seesaw, as pointed out in Ref.~\cite{Chacko:2020zze}. In this setup the composite singlet mass satisfies $M_N \sim \Lambda$, so that
\begin{equation} 
\label{eq:numass} 
 {m_\nu} \sim \mu^c \left(\frac{\lambda v_{\rm EW}}{M_N}\right)^2 
\sim \Lambda \left[C_\mu \hat{\mu}^c 
\left(\frac{\Lambda}{M_{\mathrm{UV}}}\right)^{\Delta_{2\mathrm{N^c}}-4} 
\right] 
\left[C_\lambda \hat{\lambda} \left(\frac{v_{\mathrm{EW}}}{\Lambda}\right) 
\left(\frac{\Lambda}{M_{\mathrm{UV}}}\right)^{\Delta_{\mathrm{N}}-3/2} 
\right]^2.
\end{equation}
We can further define $\varepsilon\equiv\Lambda/M_{\rm UV}$, and the expression above simplifies to
\begin{equation}
    m_\nu\sim\mathcal{O}(1) \times \Lambda\left(\frac{v_{\rm EW}}{\Lambda}\right)^2 \times\varepsilon^{\Delta_{2\mathrm{N^c}}+2\Delta_{\mathrm{N}}-7},
\end{equation}
and therefore neutrino masses are parametrically small  as long as the sum of anomalous dimensions satisfy $\Delta_{2\mathrm{N^c}}+2\Delta_{\mathrm{N}}\gtrsim 7$, though lower values would still be theoretically viable.
As pointed out in Ref.~\cite{Chacko:2020zze}, $\Delta_{2\mathrm{N^c}}+2\Delta_{\mathrm{N}}<8$ ensures no UV sensitivity to neutrino masses via the Weinberg operator.
The scaling dimensions \(\Delta_{\mathrm{N}}\) and \(\Delta_{\rm 2N^c}\) provide natural suppression of neutrino masses. For the inverse seesaw formula to be valid, we require
\(\mu^c \lesssim \Lambda\), which implies
\begin{equation}
\label{lambdabound}
\frac{\sqrt{m_\nu M_N}}{v_{\rm EW}} \lesssim \lambda \lesssim \frac{M_N}{v_{\rm EW}}\;.
\end{equation}
For multiple generations, we have pairs \(N_\alpha, N^c_\alpha\) with flavor-dependent couplings \(\lambda_{i\alpha}\). After diagonalizing the mass matrices, SM neutrinos mix with mass eigenstates \(N_\alpha\) via
\begin{equation}
U_{N_\alpha\ell_i} = \frac{\lambda_{i\alpha}\,v_{\mathrm{EW}}}{M_{N_\alpha}}\;.
\end{equation}

\section{Neutrino Dipole Moments from Compositeness}
\label{sec:dipole}

Composite neutrino models generically lead to electromagnetic dipole-moment 
operators at dimension-5 in the low-energy effective theory.  
Unlike minimal neutrino-mass constructions, where such operators arise only at 
loop level and are extremely suppressed, compositeness provides effective 
tree-level contributions that can be significantly enhanced compared to the 
simplest realizations.  
The composite structure of the heavy neutrinos \(N\) permits direct couplings to 
electromagnetic fields through operators in the conformal sector, which flow to 
observable dipole interactions after confinement.

\subsection{Transition Dipole Moments}

In the infrared effective theory below the compositeness scale \(\Lambda\), the dipole moment operator takes the form
\begin{align}
\mathcal{L}_{\rm IR} 
\supset \frac{c_D}{\Lambda^2}LHF^{\mu\nu}\sigma_{\mu\nu}N + {\rm h.c.}~,
\label{eq:dipole_IR}
\end{align}
where \(F_{\mu\nu}\) is the photon field strength tensor and \(c_D\) is a dimensionless Wilson coefficient encoding the strength of this interaction. This operator describes a transition dipole moment between the active neutrinos in \(L\) and the heavy states \(N\), generated when the Higgs obtains its vacuum expectation value. After electroweak symmetry breaking, this term yields a coupling proportional to \(c_D v_{\rm EW}/\Lambda^2\), providing a characteristic dipole interaction scale.

This operator originates from higher-dimensional interactions in the UV theory involving composite CFT operators. The transition dipole arises from
\begin{equation}
    \mathcal{L}_{\rm UV} \supset \frac{\hat{c}_D}{M_{\rm UV}^{\Delta_N+1/2}}LHF^{\mu\nu}\sigma_{\mu\nu}\mathcal{O}_{N} + {\rm h.c.}~,
\label{eq:dipole_UV_transition}
\end{equation}
where \(\hat{c}_D\) is a dimensionless coupling and \(\mathcal{O}_N\) is the fermionic composite operator introduced in Section~\ref{sec:inv_see}. The power of \(M_{\rm UV}\) in the denominator is fixed by dimensional analysis. 
The operator \(\mathcal{O}_N\) here is the same composite operator of the neutrino-mass mechanism in Eq.~\eqref{eq:LagUVss}, no additional CFT structure is introduced.  
Since \(\mathcal{O}_N\) has scaling dimension \(\Delta_N\) and 
\(LHF^{\mu\nu}\sigma_{\mu\nu}\) has canonical dimension \(9/2\), the combined 
operator has dimension \(\Delta_N + 9/2\), fixing the suppression scale in 
Eq.~\eqref{eq:dipole_UV_transition} by requiring \(M_{\rm UV}^{\Delta_N + 1/2}\) in the denominator to yield a dimension-4 Lagrangian term.

This operator represents the coupling of an electromagnetic field strength to a mixed SM-composite current, providing a portal through which the photon can probe the internal structure of the composite neutrinos.
The matching from UV to IR proceeds through the standard RG flow, with the low-energy Wilson coefficient given by
\begin{equation} 
c_D \sim C_{D} \hat{c}_D\, \left(\frac{\Lambda}{M_{\mathrm{UV}}}\right)^{\Delta_N + 1/2},
\quad
C_D = \frac{(4\pi)^{3/2-\Delta_{\mathrm{N}}}}{\Gamma(\Delta_{\mathrm{N}}-3/2)}
\sqrt{\frac{ \pi}{ (\Delta_{\mathrm{N}}-3/2)\cos (\Delta_{\mathrm{N}}\pi) }} \,.
\label{eq:Wilson_matching}
\end{equation}
The matching coefficient \(C_D\) coincides with \(C_\lambda\) of Eq.~\eqref{eq:matching_lambda}: both the mass portal of Eq.~\eqref{eq:LagUVss} and the dipole portal of Eq.~\eqref{eq:dipole_UV_transition} are built from the same fermionic operator \(\mathcal{O}_N\), so they inherit an identical normalization from its two-point function, Eq.~\eqref{eq:A_delta}.
These matching coefficients encode the nontrivial effects of the conformal dynamics, in which the anomalous dimension plays a central role. For \(\Delta_N > 3/2\), the suppression factor encoded in \(\Lambda/M_{\rm UV}\) becomes stronger, reducing the effective dipole coupling at low energies.
On the other hand, larger values of $\Delta_N$ will lead to lower scales of RH neutrinos, see Eq.~\eqref{eq:numass}.
When the scaling dimension instead approaches its free-field value, although the scale of the RH neutrinos may be higher, the
suppression of the transition magnetic moment is minimal and the dipole coupling can be relatively large.  
In this sense the setup is ``win-win’’: dimensions near the free-field limit 
enhance dipole signals, while larger anomalous dimensions amplify the effects 
relevant for neutrino-mass generation, making the two probes complementary.

After electroweak symmetry breaking, the transition dipole operator generates an effective mixing between active neutrinos and heavy neutrinos in the presence of a magnetic field. The relevant coupling strength is
\begin{equation}
d_{\mu N}\sim \frac{v_{\rm EW}}{\Lambda^2}\,\hat{c}_D C_D \left(\frac{\Lambda}{M_{\mathrm{UV}}}\right)^{\Delta_N + 1/2},
\label{eq:effdipmix}
\end{equation}
which has inverse mass dimensions and characterizes the electromagnetic interaction strength. This effective dipole coupling controls observable processes such as heavy neutrino decays \(N \to \nu \gamma\), neutrino-electron scattering with photon exchange, and other electromagnetic signatures. The key feature is that \(d_{\mu N}\) can be parametrically larger than what would arise from loop-level processes in minimal models, potentially reaching experimentally observable values.

\subsubsection{Direct Dipole Moments}

The composite nature of heavy neutrinos also permits direct magnetic moment operators of the form
\begin{equation}
\mathcal{L}_{\rm IR} \supset \frac{c_{P}}{\Lambda}\bar{N}\sigma^{\mu\nu}NF_{\mu\nu} + {\rm h.c.}~,
\label{eq:direct_dipole_IR}
\end{equation}
which arise from purely composite operators in the CFT that couple to the photon. Such operators originate in the UV theory from
\begin{equation}
    \mathcal{L}_{\rm UV} \supset \frac{\hat{c}_P}{M_{\rm UV}^{\Delta_P-2}}\mathcal{O}^{P}_{\mu\nu}F^{\mu\nu} + {\rm h.c.}~,
\label{eq:dipole_UV_direct}
\end{equation}
where \(\mathcal{O}^{P}_{\mu\nu}\) is an antisymmetric rank-2 tensor operator in the CFT with scaling dimension \(\Delta_P\).

To properly match this UV operator onto the IR theory, we must normalize \(\mathcal{O}^{P}_{\mu\nu}\) using the standard CFT prescription for antisymmetric tensor primaries. The two-point function is
\begin{equation}
\begin{split}
\int d^4 x e^{ipx}
\langle 0|T\left[ \mathcal{O}_{\mu\rho}(x) \mathcal{O}_{\nu\sigma}^\dagger(0)\right] |0 \rangle ={}&-
\frac{A_{\Delta_P}}{2i \sin \left( \Delta_P \pi \right) }
\frac{K_{\mu\nu}(p)K_{\rho\sigma}(p)-\frac{1}{4}g_{\mu\nu}g_{\rho\sigma} - (\mu\leftrightarrow \rho)}
{\left(-p^2 - i \epsilon\right)^{2 - \Delta_P} },
\end{split}
\label{eq:tensor_two_point}
\end{equation}
where 
\begin{equation}
K_{\mu\nu}= g_{\mu\nu}-\frac{p_{\mu}p_{\nu}}{p^2}
\label{eq:transverse_projector}
\end{equation}
is the transverse projector and
\begin{equation}
A_{\Delta_P} =
\frac{16 \pi^{5/2}}{\left(2 \pi \right)^{2 \Delta_P }}
\frac{\Gamma \left(\Delta_P+1/2\right)}
{\Gamma \left(\Delta_P-1\right)
\Gamma \left(2 \Delta_P\right)}\,.
\label{eq:A_Delta_P}
\end{equation}
This two-point function is well-defined for \(\Delta_P < 2\) when considering the full correlator, but for \(\Delta_P \geq 2\) UV divergences appear. The upper bound \(\Delta_P < 3\) arises from renormalization constraints. The two-point function of \(\mathcal{O}^{P}_{\mu\nu}\) develops UV divergences for \(\Delta_P \geq 2\), but these can be absorbed into counterterms provided the absorptive part remains finite. This holds for \(2 \leq \Delta_P < 3\). At \(\Delta_P = 3\), even the absorptive part develops logarithmic divergences, and beyond this threshold a tower of higher-dimension operators involving additional SM field strength insertions would be required for renormalizability. This proliferation of counterterms signals an EFT breakdown. Since \(\mathcal{O}^{P}_{\mu\nu}\) couples directly to the SM photon field strength, we require \(2 \leq \Delta_P < 3\) for a consistent EFT description with a controlled set of counterterms.

The matching from UV to IR proceeds analogously to the transition dipole case, with the low-energy Wilson coefficient given by
\begin{equation} 
c_P \sim C_{\Delta_P} \hat{c}_P\, \left(\frac{\Lambda}{M_{\mathrm{UV}}}\right)^{\Delta_P - 2},
\label{eq:Wilson_matching_direct}
\end{equation} 
where 
\begin{equation}
C_{\Delta_P} = \frac{(4\pi)^{2-\Delta_P}}{\Gamma(\Delta_P-1)} 
\sqrt{\frac{1}{\Delta_P-1}}~,
\label{eq:C_Delta_P}
\end{equation}
For \(\Delta_P > 2\), the direct dipole coupling receives stronger RG suppression than the transition dipole when \(\Delta_P > \Delta_N + 1/2\), allowing hierarchies \(c_P \ll c_D\) to naturally emerge from the composite dynamics.

\subsection{Production via Dipole Scattering}

\begin{figure}[tb]
\centering
\includegraphics[width=0.5\textwidth]{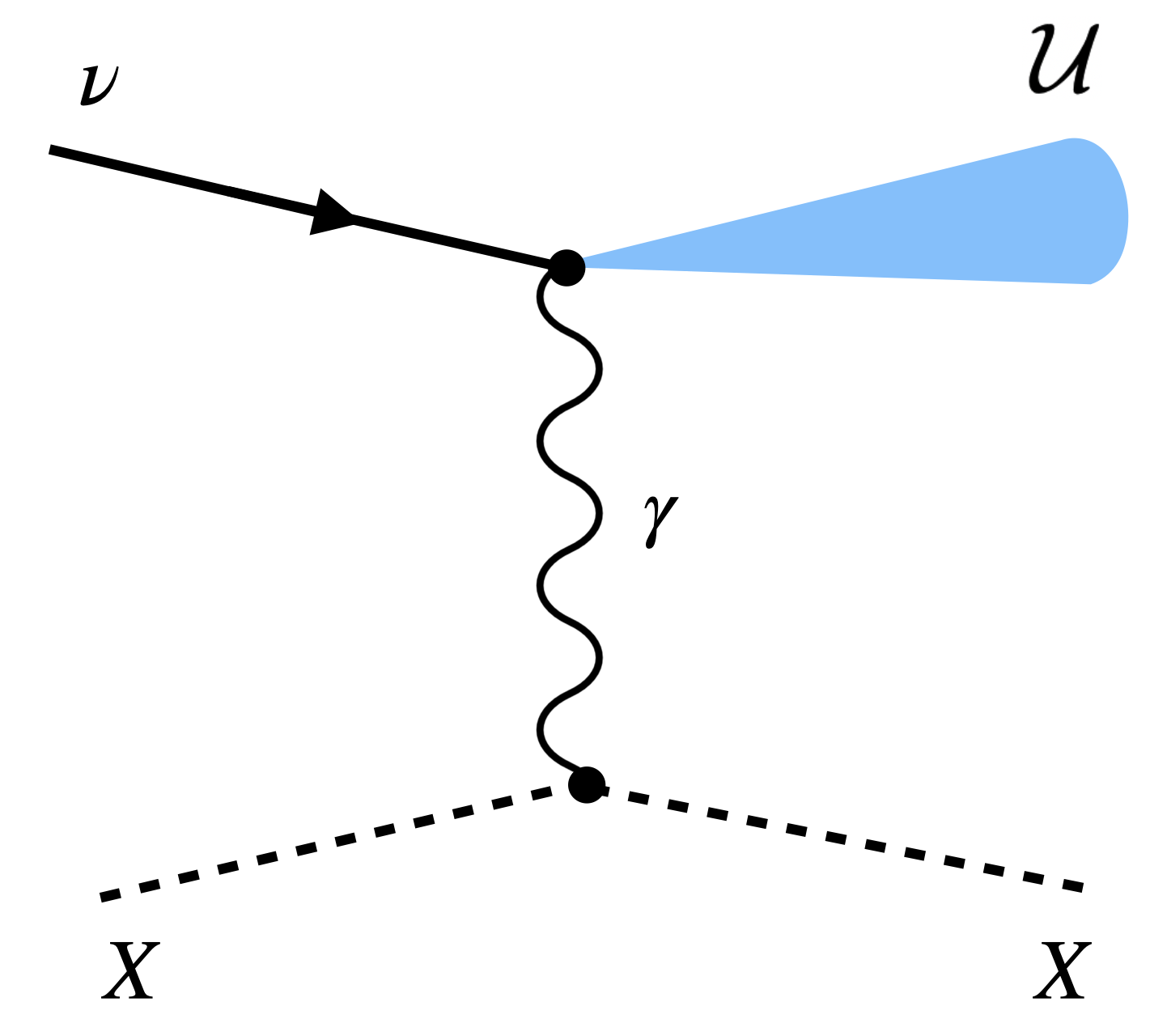}
\caption{Feynman diagram for dipole-mediated production of unparticle \(\mathcal{U}\) via the process \(\nu X \to \mathcal{U} X\).
}
\label{fig:production_diagram}
\end{figure}

We consider the upscattering process \(\nu \, X \to \mathcal{U}\, X\), where the active neutrino interacts with a target particle $X$ of charge \(Z\) through photon exchange mediated by the transition dipole operator. 
In this work, we will focus on neutrino-nucleus coherent elastic scattering.
The diagram for this process is shown in Fig.~\ref{fig:production_diagram}.
The amplitude is given by
\begin{equation}
\mathcal{M}_{\rm prod} = i \hat{c}_D e  Z F(q^2)\frac{v_{\rm EW}}{M_{\text{UV}}^{\Delta_N + 1/2}} \frac{g^{\mu\nu}}{q^2} \, q^\rho \left( \bar{u}_X \gamma_\mu u_X \right) \left( \bar{u}_{\mathcal{U}} \sigma_{\nu\rho} P_L u_\nu \right),
\label{eq:production_amplitude}
\end{equation}
where $q$ is the momentum exchange, \(t = q^2 \) is the Mandelstam variable, $F$ is a nuclear form factor~\cite{Helm:1956zz}, \(e\) is the electric charge, and \(P_L = (1-\gamma_5)/2\) is the left-handed projector. 
This amplitude encodes the coupling of the electromagnetic current of the target to the mixed neutrino-dipole vertex.

The cross-section for this process depends crucially on the phase space available to the unparticle state \(\mathcal{U}\). 
Since \(\mathcal{U}\) is a composite object emerging from the near-conformal sector, its phase space differs from that of an elementary particle. 
In the UV theory, the phase space for the composite operator \(\mathcal{O}_N\) takes the unparticle form
\begin{equation}
d\Phi_{\mathcal{O}} = \frac{A_{\Delta_{\mathrm{N}} - 1/2}}{(2\pi)^4} \, \theta(p_{\mathcal{U}}^0) \, \theta(p_{\mathcal{U}}^2 - \mu_{\text{IR}}^2) \, \left(p_{\mathcal{U}}^2 - \mu_{\text{IR}}^2\right)^{\Delta_{\mathrm{N}} - 5/2} \, d^4 p_{\mathcal{U}},
\label{eq:unparticle_phase_space}
\end{equation}
where \(\mu_{\text{IR}} \sim \Lambda\) is an infrared cutoff reflecting the scale at which conformal symmetry breaks. 
This phase space factor differs from the ordinary \(\delta(p^2 - M^2)\) constraint of an elementary particle, instead exhibiting a continuous spectral density characteristic of unparticle physics.

Integrating over $M_{\mathcal{U}}^2$ and writing $d^4p = dM_{\mathcal{U}}^2\,d^3p/(2E_{\mathcal{U}})$, the phase space becomes
\begin{equation}
d\Phi_{\mathcal{O}} = A_{\Delta_{\mathrm{N}} - 1/2} \, \frac{d^3 p_{\mathcal{U}}}{(2\pi)^3\, 2 E_{\mathcal{U}}} \, \left(M_{\mathcal{U}}^2 - \mu_{\text{IR}}^2\right)^{\Delta_{\mathrm{N}} - 5/2}\theta(M_{\mathcal{U}}^2 - \mu_{\text{IR}}^2)\,dM_{\mathcal{U}}^2\,.
\label{eq:integrated_phase_space}
\end{equation}
This expression applies to a massless composite fermion ($\mu_{\rm IR}=0$). For $\mu_{\rm IR}>0$, confinement introduces a spectral gap. 
Conservation of fermionic quantum numbers forbids states with invariant mass in the interval $(\mu_{\rm IR},\,3\mu_{\rm IR})$: the lightest admissible state has $M_{\mathcal{U}}=\mu_{\rm IR}$, that is, a single composite fermion, while the next threshold, corresponding to three-body states, opens at $M_{\mathcal{U}}=3\mu_{\rm IR}$. 
We therefore replace the spectral weight in Eq.~\eqref{eq:integrated_phase_space} by
\begin{equation}
\left(M_{\mathcal{U}}^2-\mu_{\rm IR}^2\right)^{\Delta_N-5/2}\theta(M_{\mathcal{U}}^2-\mu_{\rm IR}^2)
\;\to\;
W\,\delta(M_{\mathcal{U}}^2-\mu_{\rm IR}^2)
+\left(M_{\mathcal{U}}^2-\mu_{\rm IR}^2\right)^{\Delta_N-5/2}\theta(M_{\mathcal{U}}^2-9\mu_{\rm IR}^2)\,,
\label{eq:spectral_massive}
\end{equation}
where the residue $W$ is fixed by collapsing the integral over the single-body window $[\mu_{\rm IR}^2,\,9\mu_{\rm IR}^2]$ onto the discrete state,
\begin{equation}
W = \int_{\mu_{\rm IR}^2}^{9\mu_{\rm IR}^2}dM^2\!\left(M^2-\mu_{\rm IR}^2\right)^{\Delta_N-5/2}
= \frac{\left(8\mu_{\rm IR}^2\right)^{\Delta_N-3/2}}{\Delta_N-3/2}\,.
\label{eq:residue_W}
\end{equation}
In the limit $\mu_{\rm IR}\to 0$, $W\to 0$ for $\Delta_N>3/2$ and $\theta(M_{\mathcal{U}}^2-9\mu_{\rm IR}^2)\to\theta(M_{\mathcal{U}}^2)$, recovering Eq.~\eqref{eq:integrated_phase_space}.

The differential cross-section for producing an unparticle state with invariant mass \(M_{\mathcal{U}}\) and momentum transfer squared \(t = q^2\) is
\begin{align}\label{eq:diff_xsec_production}
\frac{d\sigma}{dM_{\mathcal{U}}^2 \, dt} &= A_{\Delta_{\mathrm{N}} - 1/2} \cdot \frac{\hat{c}_D^2 e^2 v_{\rm EW}^2 Z^2 F^2(q^2)}{32 \pi E_\nu^2 m_X^2 t^2 M_{\text{UV}}^{2\Delta_N + 1}} \left(M_{\mathcal{U}}^2 - \mu_{\text{IR}}^2\right)^{\Delta_N - 5/2}  \\
&\quad \times \left[ M_{\mathcal{U}}^2 t \left(2 m_X (2 E_\nu + m_X) + t\right) - 4 E_\nu m_X t (2 E_\nu m_X + t) - M_{\mathcal{U}}^4 (2 m_X^2 + t) \right],\nonumber
\end{align}
where \(E_\nu\) is the incident neutrino energy and \(m_X\) is the target mass. This expression exhibits the characteristic \(1/t^2\) behavior of magnetic dipole interactions from the photon propagator, modulated by the unparticle spectral density.

Neutrino--electron scattering is recovered from Eq.~\eqref{eq:diff_xsec_production} by setting $F(q^2)=1$ and replacing $m_X\to m_e$. Other interaction channels relevant at higher energies, such as quasi-elastic scattering, resonance production, and deep-inelastic scattering, require a more complete treatment of nuclear effects; we leave their inclusion to future work. A framework well-suited to this extension is provided by \textsc{Achilles}~\cite{Isaacson:2022cwh}, which is capable of interfacing BSM production cross sections with nuclear physics in neutrino scattering.

\subsection{Transition from Unparticles to Particles}
\label{subsec:part_unpart}

The spectral decomposition in Eq.~\eqref{eq:spectral_massive} resolves the unparticle into a discrete state at $M_{\mathcal{U}}=\mu_{\rm IR}$ and a continuum with $M_{\mathcal{U}}\geq 3\mu_{\rm IR}$. At confinement, each component fragments into on-shell RH neutrinos. Conservation of fermionic quantum numbers requires the final state to contain an odd number of net composite fermions. In the minimal spectrum, where $\mathcal{O}_N$ excites only the single lightest composite fermion $N$, the produced states therefore carry an odd multiplicity $n$ of $N$'s. The discrete component has $M_{\mathcal{U}}=\mu_{\rm IR}$ and produces $n=1$ exactly. For the continuum, the minimal-spectrum multiplicities are odd integers in $\{3,5,\ldots,n_{\rm max}\}$, with $n_{\rm max}=\lfloor M_{\mathcal{U}}/\mu_{\rm IR}\rfloor$ rounded to the nearest odd integer.
We caution that this odd-only counting holds only under the assumption that the lightest composite fermion $N$ is the sole resonance excited by $\mathcal{O}_N$ below $3\mu_{\rm IR}$. In a realistic theory $\mathcal{O}_N$ also excites heavier resonances with the same quantum numbers; if any such resonance lies between $2\mu_{\rm IR}$ and $3\mu_{\rm IR}$, it decays as (heavy)$\,\to 2N+\nu$, populating \emph{even} multiplicities of $N$ already above $M_{\mathcal{U}}=2\mu_{\rm IR}$~\cite{Ahmed:2025ldh}. Our analysis adopts the conservative limit in which all such resonances are heavy enough that multiple-$N$ states first appear only at $M_{\mathcal{U}}\geq 3\mu_{\rm IR}$, so the results below are unchanged; the opposite extreme, in which $\geq 2N$ states appear already above $2\mu_{\rm IR}$, would enhance the multi-photon signal and is left to future work.

The fragmentation dynamics of the continuum are not calculable from first principles, analogously to fragmentation functions in QCD. We parametrize this uncertainty through the conditional probability $P(n\,|\,M_{\mathcal{U}})$ of producing $n$ RH neutrinos from an unparticle of invariant mass $M_{\mathcal{U}}$. We adopt a phase-space-weighted model in which $P$ is proportional to the $n$-body phase-space volume near threshold,
\begin{equation}
P_{\rm psw}(n\,|\,M_{\mathcal{U}}) \;\propto\; \frac{(M_{\mathcal{U}}-n\,\mu_{\rm IR})^{2n-4}}{(n-1)!\,(n-2)!}\,, \qquad n\geq 3\,,
\label{eq:physics_mode}
\end{equation}
normalized to unity over the allowed odd multiplicities. The factorial denominators disfavor high multiplicities when $M_{\mathcal{U}}$ is close to $n\,\mu_{\rm IR}$.

\begin{figure}[tb]
  \centering
\includegraphics[width=0.4\textwidth]{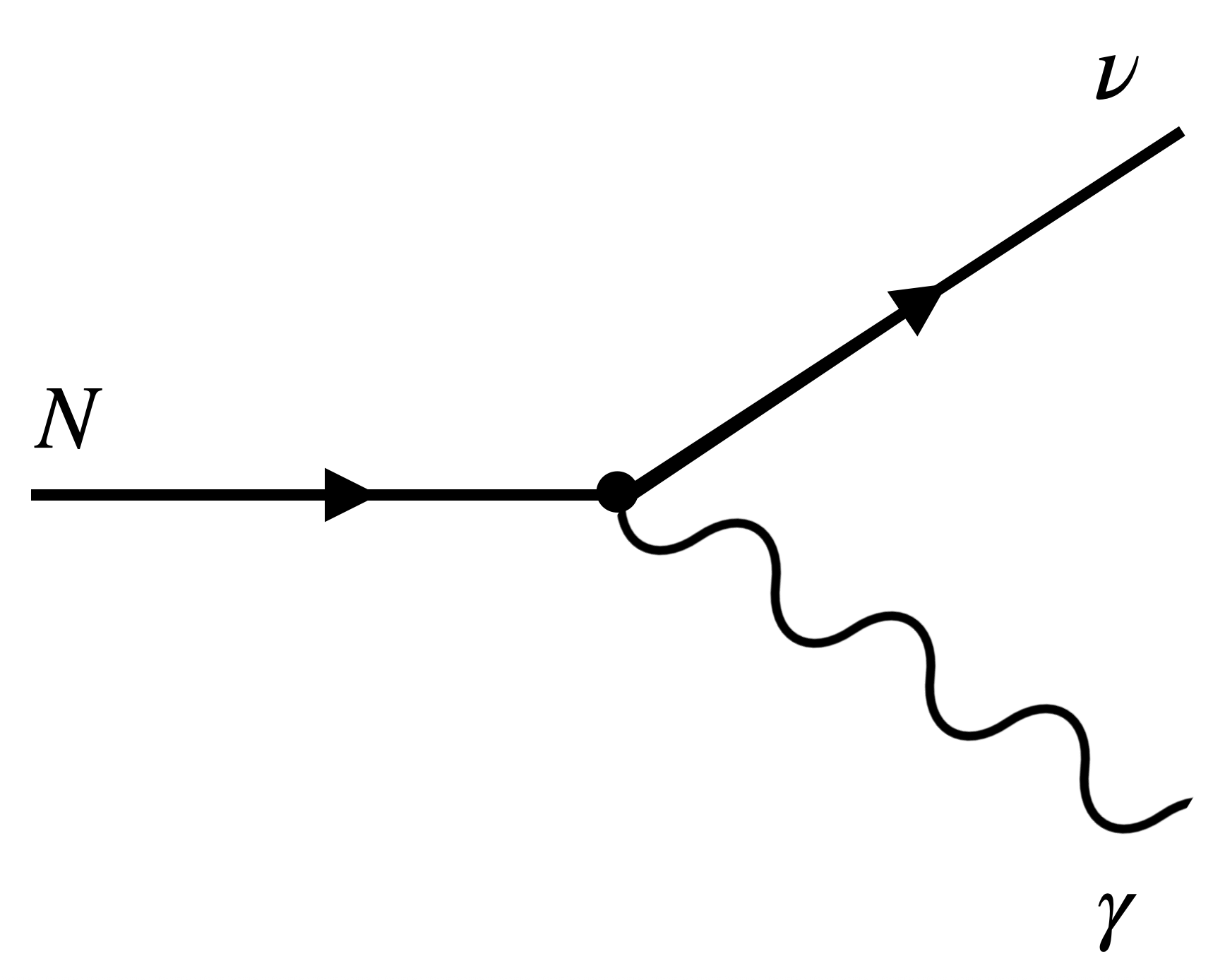}
  \caption{Feynman diagram for the radiative decay \(N \to \nu \gamma\) via the transition dipole moment operator.}
  \label{fig:decay_diagram}
\end{figure}

Once $n$ is sampled, the $n$-body phase space is generated by recursive $1\to 2$ splittings in the rest frame of the unparticle with isotropic angular distributions. Each RH neutrino then decays independently as $N\to\nu\gamma$.
Once produced, the heavy composite neutrino \(N\) decays primarily through the radiative channel \(N \to \nu \gamma\) mediated by the dipole operator. The Feynman diagram for this process is shown in Fig.~\ref{fig:decay_diagram}.
For a single heavy neutrino mode with fixed mass $M_N$, the decay amplitude is
\begin{equation}
\mathcal{M}_{\rm decay} 
= i\, d_{\mu N}\, \bar{u}(p_N)\, \sigma^{\mu\nu} P_L\, u(p_\nu)\, 
\epsilon_\mu(q)\, q_\nu,
\label{eq:decay_amplitude}
\end{equation}
where \(d_{\mu N}\) is the effective dipole coupling for that
mass eigenstate from Eq.~\eqref{eq:effdipmix}, \(q\) is the photon momentum, \(\epsilon_\mu(q)\) is the photon polarization vector, and \(P_L\) is the left-handed projector.
It is then trivial to obtain the RH neutrino total width,
\begin{equation}
\Gamma(N \to \nu \gamma) = \frac{d_{\mu N}^2 M_N^3}{4\pi}.
\label{eq:total_decay_width_simple}
\end{equation}

For typical benchmark parameters with \(M_N \sim 100\)--\(500\) MeV and dipole couplings \(d_{\mu N} \sim 10^{-7}\)--\(10^{-8}\,{\rm GeV}^{-1}\), the proper decay length is \(c\tau = \hbar c/\Gamma \sim 10\)--\(10^3\) m. In the lab frame with typical boosts \(\gamma \sim 2\)--\(5\), this yields decay distances of order 20 m to several km, meaning \(N\) decays in-flight within or beyond typical neutrino detector volumes (\(\sim 10\) m scale). 
Only for the largest dipole couplings in our range (\(d_{\mu N} \sim 10^{-6}\,{\rm GeV}^{-1}\)) and highest masses (\(M_N \gtrsim 0.5\) GeV) do the decay lengths approach prompt scales of order cm to m.

\section{Phenomenology}
\label{sec:pheno}

We now turn to the phenomenology of the transition dipole portal: the parameter space probed by current experiments, the resulting bounds, and the complementary cosmological, astrophysical, and collider constraints. Table~\ref{tab:newbenchmarks} collects six benchmark scenarios, assuming a single active neutrino family, spanning compositeness scales \(\Lambda \sim 50\)--\(500\) MeV, UV cutoffs \(M_{\rm UV} \sim 0.5\)--\(10\) TeV, and scaling dimensions \(\Delta_N \in [1.6, 2.4]\). The UV couplings \(\hat{c}_D \lesssim 1\) are natural for a strongly-coupled sector. The active--sterile mixing \(|U|^2 \sim 10^{-16}\)--\(10^{-6}\) is naturally suppressed by the composite dynamics.

\begin{table}[t]
  \centering
  \resizebox{\textwidth}{!}{
  \begin{tabular}{c|cccccccccc}
    \hline\hline
    Benchmark
    & \(\hat{\lambda}\)
    & \(\hat{\mu}^c\)
    & \(\hat{c}_D\)
    & \(\Delta_N\)
    & \(\Delta_{2N^c}\)
    & \(\Lambda\) [MeV]
    & \(M_{\rm UV}\) [TeV]
    & \(m_\nu\) [eV]
    & \(|U|^2\)
    & \(d_{\mu N}\) [GeV\(^{-1}\)] \\
    \hline
    A
    & \(2.64\times10^{-4}\)
    & \(8.92\times10^{-6}\)
    & 0.5
    & 2.4
    & 3.00
    & 100
    & 0.5
    & \(1.24\times10^{-3}\)
    & \(4.82\times10^{-9}\)
    & \(5.25\times10^{-8}\) \\
    B
    & \(1.94\times10^{-4}\)
    & \(4.18\times10^{-6}\)
    & \(0.22\)
    & 2.0
    & 3.15
    & 500
    & 5.0
    & \(1.32\times10^{-2}\)
    & \(7.25\times10^{-8}\)
    & \(6.13\times10^{-9}\) \\ \hline
    C
    & \(6.39\times10^{-5}\)
    & \(1.44\times10^{-6}\)
    & \(0.0624\)
    & 2.0
    & 3.02
    & 50
    & 0.5
    & \(2.53\times10^{-2}\)
    & \(7.88\times10^{-7}\)
    & \(1.73\times10^{-7}\) \\
    D
    & \(3.20\times10^{-5}\)
    & \(1.28\times10^{-5}\)
    & \(0.294\)
    & 2.0
    & 3.20
    & 100
    & 1.0
    & \(3.10\times10^{-3}\)
    & \(4.94\times10^{-8}\)
    & \(2.04\times10^{-7}\) \\ 
    E
    & \(1.00\times10^{-8}\)
    & \(7.56\times10^{-4}\)
    & 1.0
    & 2.0
    & 2.39
    & 100
    & 10
    & \(1.51\times10^{-3}\)
    & \(4.82\times10^{-16}\)
    & \(2.20\times10^{-9}\) \\
    F
    & \(2.82\times10^{-7}\)
    & \(1.00\times10^{-8}\)
    & \(3.58\times10^{-3}\)
    & 1.6
    & 2.57
    & 100
    & 1.0
    & \(2.75\times10^{-3}\)
    & \(2.58\times10^{-8}\)
    & \(2.04\times10^{-7}\) \\
    \hline\hline
  \end{tabular}
  }
  \caption{
  Benchmark parameter sets used in Fig.~\ref{fig:dipole_bounds}.
  Benchmarks A and B correspond to the reference scenarios with
  \(\Delta_N=2.4\) and \(\Delta_N=2.0\), respectively.
  Benchmarks C and D illustrate the dependence on the composite neutrino mass
  \(m_N\), benchmark E probes sensitivity to the UV cutoff scale \(M_{\rm UV}\),
  and benchmark F demonstrates the impact of a reduced scaling dimension
  \(\Delta_N\).
  The table lists the UV couplings \(\hat{\lambda}\),
  \(\hat{\mu}^c\), and \(\hat{c}_D\),
  the scaling dimensions \(\Delta_N\) (fermionic composite operator) and \(\Delta_{2N^c}\) (scalar bilinear composite, governing neutrino mass suppression),
  the compositeness scale \(\Lambda \simeq m_N\),
  the UV cutoff \(M_{\rm UV}\),
  the induced neutrino mass \(m_\nu\),
  the active--sterile mixing \(|U|^2\),
  and the effective dipole coupling \(d_{\mu N}\).
  }
  \label{tab:newbenchmarks}
\end{table}

\begin{figure}[t]
  \centering
  \includegraphics[width=0.49\textwidth]{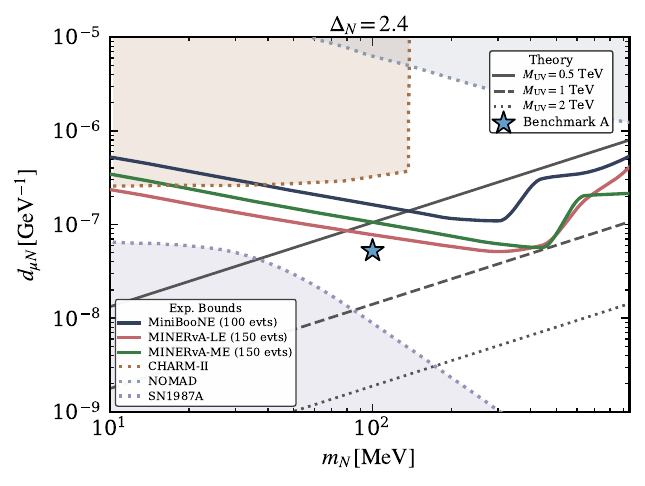}
  \includegraphics[width=0.49\textwidth]{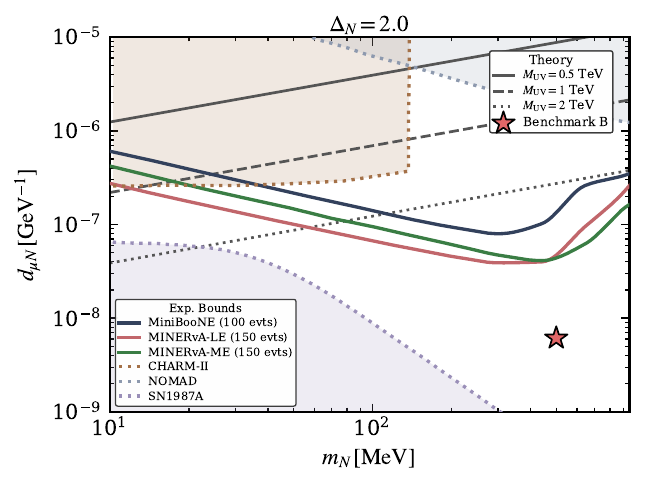}
  \caption{
  Effective dipole coupling \(d_{\mu N}\) as a function of the heavy neutrino mass
  \(m_N \sim \Lambda\), for \(\Delta_N = 2.4\) (left) and \(\Delta_N = 2.0\) (right).
  In each panel, the gray curves show the theoretical upper envelope obtained by
  setting \(\hat{c}_D = 1\), for \(M_{\rm UV} = 0.5,\,1,\) and \(2~\mathrm{TeV}\)
  (solid, dashed, dotted respectively).
  The MiniBooNE and MINERvA estimated sensitivities (solid lines) are computed
  within our model from the simulated production-and-decay event yields,
  corresponding to \(100\) signal events for MiniBooNE and \(150\) for MINERvA;
  for MINERvA we show both the NuMI low-energy (LE) and medium-energy (ME) beam
  configurations. The CHARM-II and NOMAD shaded regions are model-independent transition-dipole limits, from neutrino–electron scattering and Primakoff single-photon production respectively, and the SN1987A bound is taken from the literature.
  The star markers denote benchmark points from Table~\ref{tab:newbenchmarks},
  with benchmark~A (B) corresponding to the \(\Delta_N=2.4\) (\(\Delta_N=2.0\))
  scenario.
  }
  \label{fig:dipole_bounds}
\end{figure}

The theoretical predictions and current experimental constraints on the dipole coupling \(d_{\mu N}\) are summarized in Figure~\ref{fig:dipole_bounds}. The gray curves show the maximum predicted \(d_{\mu N}\) as a function of \(m_N \sim \Lambda\) for \(M_{\rm UV} = 0.5\), \(1\), and \(2\) TeV (solid, dashed, dotted), obtained by setting \(\hat{c}_D = 1\); this represents an approximate theoretical upper bound on the dipole coupling for a given compositeness scale and UV cutoff, with smaller \(\hat{c}_D\) shifting the prediction downward.

Two features of Figure~\ref{fig:dipole_bounds} merit emphasis. First, the model naturally predicts dipole couplings in the range \(10^{-8}\)--\(10^{-6}\,{\rm GeV}^{-1}\) for compositeness scales between 10 MeV and 1 GeV, significantly enhanced compared to standard loop-level expectations in minimal models. Second, the variation with \(M_{\rm UV}\) demonstrates that lower UV cutoffs lead to larger dipole couplings for a given compositeness scale. The experimental, cosmological, and astrophysical results overlaid in this figure are discussed in turn below, after we describe the event simulation used to derive the MiniBooNE and MINERvA sensitivities.

\subsection{Event Simulation}
\label{subsec:simulation}

To compare the model against neutrino-beam data we compute the expected number of signal events through a dedicated event-level simulation of the full chain \(\nu X \to \mathcal{U} X \to \nu + n\gamma\, X\). The expected yield at a given experiment takes the form
\begin{equation}
N_{\rm sig} = N_{\rm tgt} T \int d\vec{x}\,dE_\nu\,\frac{d\phi_\nu}{dE_\nu}\,
\sigma(E_\nu;\,\mu_{\rm IR},d_{\mu N})\,
    \epsilon_{\rm dec}(E_\nu, \vec{x};\,\mu_{\rm IR},d_{\mu N})\,\epsilon_{\rm det}\,,
\label{eq:Nsig}
\end{equation}
where \(d\phi_\nu/dE_\nu\) is the incident neutrino flux, \(T\) and \(N_{\rm tgt}\) are the exposure and number of target scattering centers in the production region, \(\sigma\) is the coherent production cross section of Eq.~\eqref{eq:diff_xsec_production} integrated over \(M_{\mathcal{U}}^2\) and \(t\), \(\epsilon_{\rm dec}\) is the probability that after an unparticle is produced at \(\vec{x}\) at least one daughter \(N\) decays radiatively inside the fiducial volume, and \(\epsilon_{\rm det}\) is the photon reconstruction and selection efficiency. The unparticle can be produced anywhere along the beam line, not only inside the detector, so we sample the production point \(\vec{x}\) over a sequence of three contiguous slabs along the beam axis, that is, an extended upstream target of dirt or rock, a short air gap in front of the detector, and the detector medium itself, and retain only events whose daughter \(N\) decays within the fiducial volume. Because we lack the experimental systematics and selection details needed for a full search, we model the detector response explicitly rather than through a constant reconstruction efficiency: we set \(\epsilon_{\rm det}=1\) and let detector effects enter via a photon detection threshold \(E_\gamma>20\)~MeV, a Gaussian angular smearing of \(2^\circ\), and an electromagnetic energy resolution \(\sigma_E/E = 10\%/\sqrt{E/{\rm GeV}}\). 
The resulting MiniBooNE and MINERvA curves are therefore estimated sensitivities rather than rigorous exclusion limits; see Ref.~\cite{Kamp:2022bpt} for a detailed study of the elementary transition dipole scenario.

For MiniBooNE we use the Booster Neutrino Beam \(\nu_\mu\) flux in neutrino mode~\cite{MiniBooNE:2008hfu} and an exposure of \(1.875\times10^{21}\) POT, the full neutrino-mode dataset of Ref.~\cite{MiniBooNE:2020pnu}. The production volume comprises \(\approx 400\)~m of dirt spanning \(50\)--\(446\)~m from the target, modeled as a silicon proxy with \(Z=14\), \(A=28\) and atomic density \(3.87\times10^{22}~{\rm cm}^{-3}\); a \(4\)~m air gap in front of the cryostat, a nitrogen proxy with \(Z=7\), \(A=14\) and density \(5.27\times10^{19}~{\rm cm}^{-3}\); and the mineral-oil detector approximated as carbon with \(n_C = 4.27\times10^{22}~{\rm cm}^{-3}\). The fiducial decay volume is a box of transverse half-width \(4.05\)~m spanning \(450\)--\(458\)~m from the target, containing \(N_{\rm tgt}=2.3\times10^{31}\) carbon nuclei.

For MINERvA we consider both NuMI beam configurations: the low-energy (LE) flux in forward-horn-current \(\nu_\mu\) mode~\cite{MINERvA:2016iqn} with an exposure of \(4\times10^{20}\) POT, and the medium-energy (ME) flux~\cite{MINERvA:2019hhc} with \(1.16\times10^{21}\) POT. The two beams share the same detector and production geometry, differing only in the flux shape and exposure. The production volume follows the same three-slab construction as for MiniBooNE: \(\approx 230\)~m of dolomite rock spanning \(785\)--\(1015\)~m from the target, a silicon proxy with density \(3.87\times10^{22}~{\rm cm}^{-3}\); \(\approx 20\)~m of MINOS-hall air, a nitrogen proxy with density \(5.27\times10^{19}~{\rm cm}^{-3}\); and the active scintillator tracker, polystyrene CH modeled as carbon with \(n_C=4.83\times10^{22}~{\rm cm}^{-3}\), at \(1.035\)~km from the target~\cite{MINERvA:2013zvz,Adamson:2015dkw}. The fiducial volume is a rectangular approximation to the \(85\)~cm-apothem tracker hexagon, with transverse half-width \(0.80\)~m, \(2\)~m long, and \(N_{\rm tgt}=2.5\times10^{29}\) carbon nuclei.

For each sampled production event, the unparticle invariant mass \(M_{\mathcal{U}}\) is drawn from the spectral density of Eq.~\eqref{eq:spectral_massive}, fragmented into \(n\) heavy neutrinos according to the phase-space-weighted multiplicity model of Section~\ref{subsec:part_unpart}, and each \(N\) is decayed isotropically as \(N\to\nu\gamma\) in its rest frame and boosted to the lab frame. The decay acceptance \(\epsilon_{\rm dec}\) is governed by the lab-frame decay length \(\gamma\beta\,c\tau\), with the proper lifetime fixed by the radiative width \(\tau = \hbar/\Gamma\) and \(\Gamma\) from Eq.~\eqref{eq:total_decay_width_simple}; for the parameters of interest \(c\tau\) ranges from tens of centimetres to several kilometres, so the in-volume decay probability is a strong function of \((M_N, d_{\mu N})\) and the beam energy.

Because the production rate scales as \(\sigma \propto \hat{c}_D^2/M_{\rm UV}^{2\Delta_N+1} \propto d_{\mu N}^2\) [cf.\ Eqs.~\eqref{eq:effdipmix} and~\eqref{eq:diff_xsec_production}], the event yield at fixed \(\Delta_N\) and \(M_N\sim\mu_{\rm IR}\) depends on the UV parameters only through the effective coupling \(d_{\mu N}\); the resulting contours can therefore be presented directly in the \((m_N,\,d_{\mu N})\) plane.

\subsection{Neutrino Experiments}
\label{subsec:neutrino}

At the compositeness scales of interest, the strongest sensitivity to the transition dipole comes from neutrino-beam and neutrino-scattering experiments, summarized in Figure~\ref{fig:dipole_bounds}. MiniBooNE and MINERvA dominate the reach in the region of interest, so we simulate them in detail following Section~\ref{subsec:simulation}, while CHARM-II and NOMAD enter through the simpler bounds described below. Rather than rescaling the published single-photon limits, we use the simulated production-and-decay yields \(N_{\rm sig}\) of Section~\ref{subsec:simulation} and define the estimated sensitivity by the criterion \(N_{\rm sig} > N_{\rm lim}\), with \(N_{\rm lim}=100\) signal events for MiniBooNE and \(N_{\rm lim}=150\) for MINERvA. Because these yields are computed within the model, they already fold in the composite production cross section and decay kinematics; as we do not include the experimental systematics or selection efficiencies of a dedicated search, the resulting MiniBooNE and MINERvA curves (solid lines in Figure~\ref{fig:dipole_bounds}) should be read as estimated sensitivities rather than exclusion limits.

MiniBooNE observed an excess of low-energy electron-like events~\cite{MiniBooNE:2018esg, MiniBooNE:2020pnu} that could potentially be explained by heavy neutrino production and subsequent decays producing electromagnetic showers~\cite{Bertuzzo:2018itn, Ballett:2018ynz, Arguelles:2018mtc, Ballett:2019pyw, Dutta:2020scq, Abdullahi:2020nyr, Dutta:2021cip, Arguelles:2021dqn}, though recent MicroBooNE results put tension on such interpretations~\cite{MicroBooNE:2025ovj, MicroBooNE:2025rsd, MicroBooNE:2025cak, MicroBooNE:2025khi}. Our choice \(N_{\rm lim}=100\) loosely maps onto the experiment's sensitivity to electron-like events, of which it observed an excess of about \(450\)~\cite{MiniBooNE:2020pnu}.

MINERvA has searched for single-photon events in both neutrino and antineutrino beams~\cite{MINERvA:2022vmb}. A dedicated search for this signature in the composite model, with proper event selection and reconstruction, is beyond the scope of this paper; we instead set \(N_{\rm lim}=150\), which roughly corresponds to the large neutral-current \(\pi^0\) background in their \(\nu\)--\(e\) scattering analysis~\cite{MINERvA:2015nqi}, and apply the same threshold to both the LE and ME beam configurations.

CHARM-II measured the cross section for \(\nu_\mu e \to \nu_\mu e\) elastic scattering~\cite{CHARM-II:1994dzw}. A transition dipole would add a magnetic-moment-like electromagnetic contribution to this process, \(\nu_\mu e \to N e\), with the same single-electron-recoil signature. 
We compute this dipole contribution using the same coherent cross section as for production [Eq.~\eqref{eq:diff_xsec_production}] evaluated on an electron target (\(m_{\rm tgt}=m_e\), \(Z=1\), unit form factor), at the CHARM-II beam energy \(E_\nu\approx24\)~GeV with a recoil cut \(E_R>5\)~GeV, and require it not to exceed the experimental precision on the measured cross section, which we take to be \(\sim 40\%\). The bound applies as long as the heavy neutrino is light enough to be produced at this recoil, \(M_N\lesssim140\)~MeV; the resulting upper limit on \(d_{\mu N}\) is the shaded CHARM-II region in Figure~\ref{fig:dipole_bounds}.

NOMAD searched for single forward photons from \(\nu Z \to \nu Z\gamma\) (Primakoff conversion followed by radiative decay). Rather than re-simulating this process in the composite model, we adopt the published limit on the \emph{elementary} transition dipole moment obtained from this search~\cite{Gninenko:1998nn} and map it directly onto the \(d_{\mu N}\) plane. Since this treats the heavy neutrino as an elementary state carrying a magnetic moment rather than as the composite operator considered here, the NOMAD curve should be regarded as approximate. It constrains the dipole coupling for \(M_N \sim 10\)--\(100\) MeV and is shown as the shaded NOMAD region in Figure~\ref{fig:dipole_bounds}.

\subsection{Cosmological and Astrophysical Constraints}
\label{subsec:cosmo}

Beyond laboratory searches, MeV-scale states that couple to photons are subject to cosmological and astrophysical constraints. We find that the benchmark scenarios are comfortably consistent with the former and that the latter, while relevant, are already displayed in Figure~\ref{fig:dipole_bounds}.

\paragraph{Big-Bang nucleosynthesis and the CMB.}
A long-lived MeV state decaying electromagnetically after the onset of nucleosynthesis (\(t\sim1\)~s) could dissociate light nuclei or distort the CMB blackbody spectrum~\cite{Cyburt:2015mya,Aghanim:2018eyx}. In the present model the heavy neutrinos decay radiatively with width \(\Gamma = d_{\mu N}^2 M_N^3/4\pi\) [Eq.~\eqref{eq:total_decay_width_simple}], giving proper lifetimes
\begin{equation}
\tau = \frac{4\pi}{d_{\mu N}^2 M_N^3}
\;\approx\; 8\times10^{-7}\,{\rm s}\,
\left(\frac{10^{-7}\,{\rm GeV}^{-1}}{d_{\mu N}}\right)^{2}
\left(\frac{100\,{\rm MeV}}{M_N}\right)^{3}.
\end{equation}
For every benchmark in Table~\ref{tab:newbenchmarks} this yields \(\tau \lesssim {\rm few}\times10^{-3}\)~s, ranging from \(\sim2\times10^{-7}\)~s for the most strongly coupled benchmarks (D and~F) to \(\sim2\times10^{-3}\)~s for the weakly coupled benchmark~E, so all decays happen well before nucleosynthesis.
The scenario is therefore unconstrained by BBN and CMB spectral-distortion bounds across the parameter space considered.

\paragraph{SN1987A.}
The dipole portal is constrained by the energetics of SN1987A: heavy neutrinos produced in the proto-neutron-star core can carry away energy and shorten the observed neutrino burst~\cite{Magill:2018jla}. 
This bound excludes an intermediate range of couplings, bounded from below by the requirement of efficient production and from above by re-absorption (trapping) of the heavy states within the core, and is shown as the shaded SN1987A region in Figure~\ref{fig:dipole_bounds}.
Note that we are presenting the constraints from Ref.~\cite{Magill:2018jla}, which were derived for the elementary transition dipole moment scenario, though we do not expect major changes due to the composite nature of the right-handed neutrinos studied here.
We display the trapping upper edge of the excluded band, which is the boundary relevant to the parameter space probed by terrestrial experiments; the production edge lies at lower couplings. 
The SN1987A constraint is most important at the lightest masses, where it complements the laboratory bounds, and does not exclude the benchmark points.

\subsection{Other Probes}
\label{subsec:colliders}

Because the direct dipole operator of Eqs.~\eqref{eq:direct_dipole_IR} and~\eqref{eq:dipole_UV_direct} couples two heavy states to a photon, it is probed by a range of facilities outside the short-baseline neutrino program; we briefly survey these here.

The cleanest handles are at colliders, through pair production \(e^+e^- \to N\bar{N}\gamma\) at LEP and \(pp \to N\bar{N}\gamma\) at the LHC. The strongest current constraints on the sterile-to-sterile dipole coupling come from LEP monophoton-plus-missing-energy searches~\cite{Magill:2018jla}, which bound \(|c_P/\Lambda| \lesssim 10^{-8}\)--\(10^{-9}\,{\rm GeV}^{-1}\) for sterile neutrino masses in the range \(10\)--\(100\) GeV; the high-luminosity LHC is projected to improve this by roughly two orders of magnitude through long-lived-particle searches with non-pointing photons~\cite{Barducci:2023hzo}. These searches dominate for \(M_N\) at or above the electroweak scale.

For lighter masses, neutrino detectors in the far-forward direction at the LHC offer complementary reach. FASER\(\nu\) and the proposed Forward Physics Facility~\cite{Feng:2022inv} are immersed in an intense, highly boosted, and low-background neutrino flux, where the direct dipole induces up-scattering and pair production of heavy neutrinos whose subsequent radiative decays appear as displaced or collimated electromagnetic deposits; rare meson decays provide a further probe of the sub-electroweak window.

Neutrino telescopes add a qualitatively distinct signature. At the TeV--PeV energies seen by IceCube~\cite{IceCube:2016zyt} and KM3NeT~\cite{KM3Net:2016zxf}, a heavy neutrino produced through the dipole portal is strongly boosted, so its radiative decay length stretches to detector-resolvable scales. Since the direct operator connects heavy state to heavy state, an up-scattered \(N\) can radiate further \(N \to N'\gamma\) transitions, or be produced as an \(N\bar{N}\) pair, yielding a chain of spatially separated cascades within a single event---a multi-shower topology that generalizes the double-cascade signature already used to search for new physics in IceCube~\cite{Coloma:2017ppo}. The large boost and low background make such a morphology a clean discriminator.

These probes are complementary to the neutrino-beam phenomenology that is the focus of this work. Our benchmark scenarios implicitly assume \(\hat{c}_P \ll \hat{c}_D\), so that existing LEP constraints on the direct dipole are satisfied while the transition dipole remains in the observable range for neutrino experiments; this hierarchy arises naturally when \(\Delta_P \gg \Delta_N + 1/2\), since \(c_P\) then receives stronger RG suppression [cf.\ Eq.~\eqref{eq:Wilson_matching_direct}]. A comprehensive study of the direct-dipole signatures, including updated LHC constraints and the discovery potential at forward facilities, is left to future work.

\section{Kinematic Signatures}
\label{sec:sign}

Using the production and decay formalism and the event-level simulation of Sections~\ref{subsec:part_unpart} and~\ref{subsec:simulation}, we now examine the observable distributions. The combined process \(\nu X \to N X \to \nu \gamma X\) yields a distinctive signature of electromagnetic energy deposits recoiling against the missing momentum carried by the final-state active neutrino. Beyond setting the event rate that enters the bounds of Section~\ref{sec:pheno}, the shapes of these distributions carry information about the underlying composite dynamics: the heavy-neutrino mass \(M_N\sim\Lambda\), the production kinematics, and the multiplicity structure of the fragmentation. It is these shapes that distinguish the signal both from Standard Model backgrounds and from competing new-physics explanations of the short-baseline excesses. We take MiniBooNE as the concrete setting, since it defines the rate context of Section~\ref{sec:pheno} and the low-energy excess, and comment throughout on how liquid-argon time-projection chambers (LArTPCs) sharpen each handle.

Cherenkov detectors such as MiniBooNE measure the integrated electromagnetic energy of a shower but cannot, on an event-by-event basis, distinguish a single photon from an electron. LArTPCs, with their fine-grained calorimetry and tracking, can do so to a useful degree. The primary handle is the ionization density at the start of the shower: a photon converts to an \(e^+e^-\) pair and deposits roughly twice the minimum-ionizing \(dE/dx\) over the first few centimetres, whereas a primary electron deposits the single-track minimum-ionizing value~\cite{MicroBooNE:2025cak}. 
A second handle commonly used in single-photon searches, the displacement of the shower start from the neutrino interaction vertex, the so-called conversion gap, could also be leveraged for such signals.
Nevertheless, since we only consider coherent neutrino-nucleus scattering, in which there is no visible hadronic recoil, this handle is unavailable.
A detailed, experiment-specific reconstruction at MicroBooNE, SBND, ICARUS and DUNE is beyond our scope and is left to future work.

In the remainder of this section we present the photon multiplicity, energy, and leading-photon angular distributions, together with the \(E\theta^2\) discriminant against neutrino--electron scattering, for benchmarks~A and~B.

\subsection{Photon Multiplicity}

\begin{figure}[tb]
  \centering 
  \includegraphics[width=0.49\textwidth]{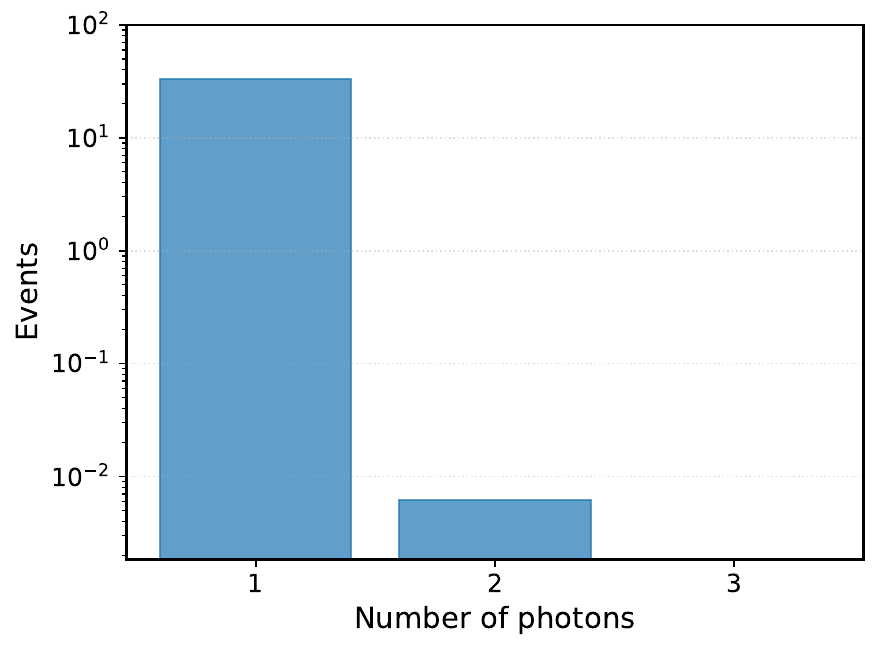}
  \includegraphics[width=0.49\textwidth]{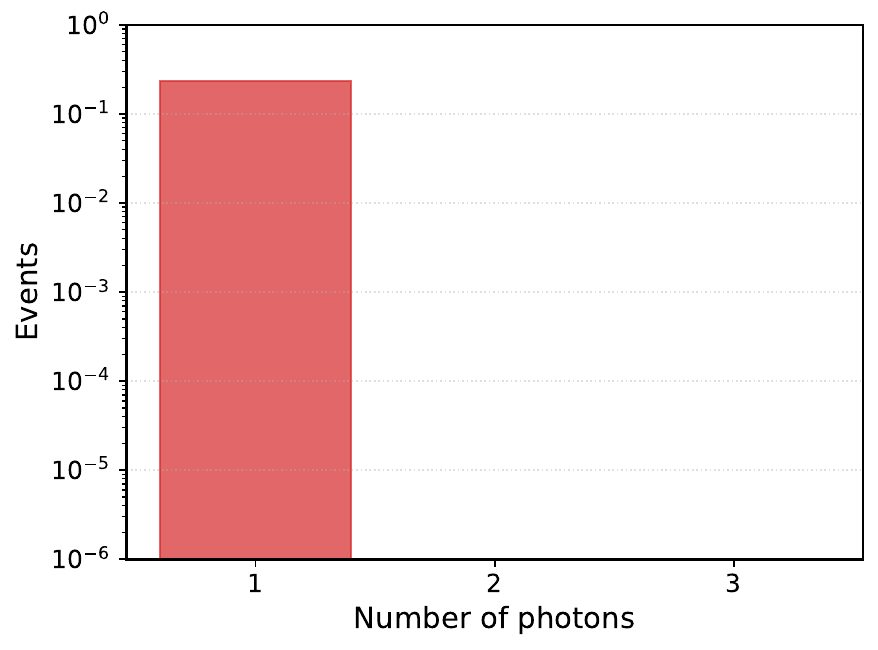}
  \caption{Photon multiplicity distribution for composite heavy neutrino decays $N \to \nu + n\gamma$ for benchmark~A (left) and benchmark~B (right).}
  \label{fig:photon_multiplicity}
\end{figure}

The photon multiplicity distribution is shown in Figure~\ref{fig:photon_multiplicity} for benchmark scenarios A and B. Although the composite production can in principle fragment the up-scattered state into several heavy neutrinos, each decaying as \(N\to\nu\gamma\), for both benchmarks the observable signal is overwhelmingly single-photon, with a multi-photon fraction below \(0.1\%\). In the conservative minimal-spectrum limit adopted here (Section~\ref{subsec:part_unpart}), fermion-number conservation restricts the \emph{produced} state to an odd number of heavy neutrinos, so the first multi-neutrino configuration is the three-\(N\) state, which requires an unparticle invariant mass \(M_{\mathcal{U}} \geq 3\mu_{\rm IR}\) to access the multi-body continuum of Eq.~\eqref{eq:spectral_massive}. Because each heavy neutrino decays inside the fiducial volume only with a small probability, however, the leading \emph{observed} correction is the two-photon final state---two of the three produced neutrinos decaying within the detector while the third escapes---as seen in Figure~\ref{fig:photon_multiplicity}; heavier composite resonances could populate even produced multiplicities already above \(2\mu_{\rm IR}\), only enhancing the multi-photon signal. The strong single-photon dominance has a two-fold origin, depending on the heavy-neutrino mass scale.

Recall that the heavy neutrino decays radiatively with width [Eq.~\eqref{eq:total_decay_width_simple}]
\begin{equation}
\Gamma(N\to\nu\gamma)=\frac{d_{\mu N}^2 M_N^3}{4\pi}\,,
\qquad
c\tau=\frac{4\pi\,\hbar c}{d_{\mu N}^2 M_N^3}\,,
\end{equation}
so the proper decay length falls steeply with mass, while the dipole coupling allowed by existing constraints shrinks as the bounds tighten toward heavier \(M_N\). The interplay of these two trends controls the multiplicity in the two regimes below.

For light heavy neutrinos, around the \(10\)~MeV scale, the experimental bounds are weakest and the largest dipole couplings are allowed, yet the \(M_N^3\) suppression of the width still makes these states long-lived, with decay lengths far exceeding the detector size. The probability that a given \(N\) decays inside the fiducial volume is then small and scales linearly with the detector length, \(P_{\rm dec}\simeq L_{\rm det}/(\gamma\beta\,c\tau)\ll1\). 
Because the long lifetime lets the heavy neutrinos survive to reach the detector from far upstream, and because the dirt presents far more target material than the detector itself, production is dominated by the upstream dirt. 
Besides, even when several \(N\) are produced together the probability that two or more of them decay within the fiducial volume scales as \(P_{\rm dec}^2\), and is therefore suppressed quadratically relative to the already small single-decay probability. 

For heavier states, around the \(500\)~MeV scale, the \(M_N^3\) growth of the width could make the decays prompt.
Nevertheless, the Booster Neutrino Beam does not offer enough energy for multi-\(N\) production, as the flux peaks near \(E_\nu\sim800\)~MeV and is strongly suppressed beyond \(\sim2\)~GeV.
The multi-body continuum of Eq.~\eqref{eq:spectral_massive} is rarely populated and the up-scattered state is almost always the discrete single-\(N\) component at \(M_{\mathcal{U}}=\mu_{\rm IR}\).

\subsection{Photon Energy Spectrum}

\begin{figure}[tb]
  \centering
  \includegraphics[width=0.49\textwidth]{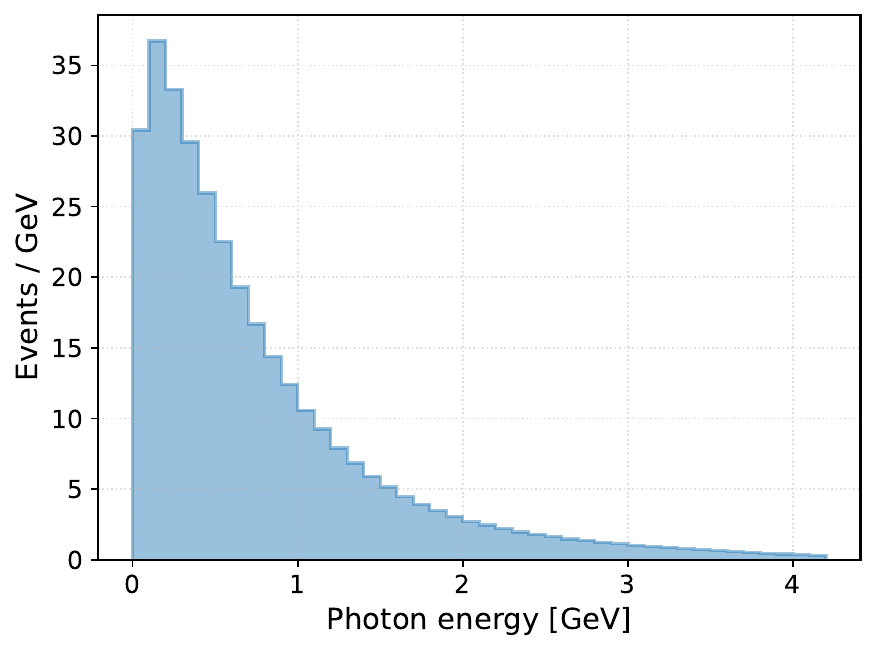}
  \includegraphics[width=0.49\textwidth]{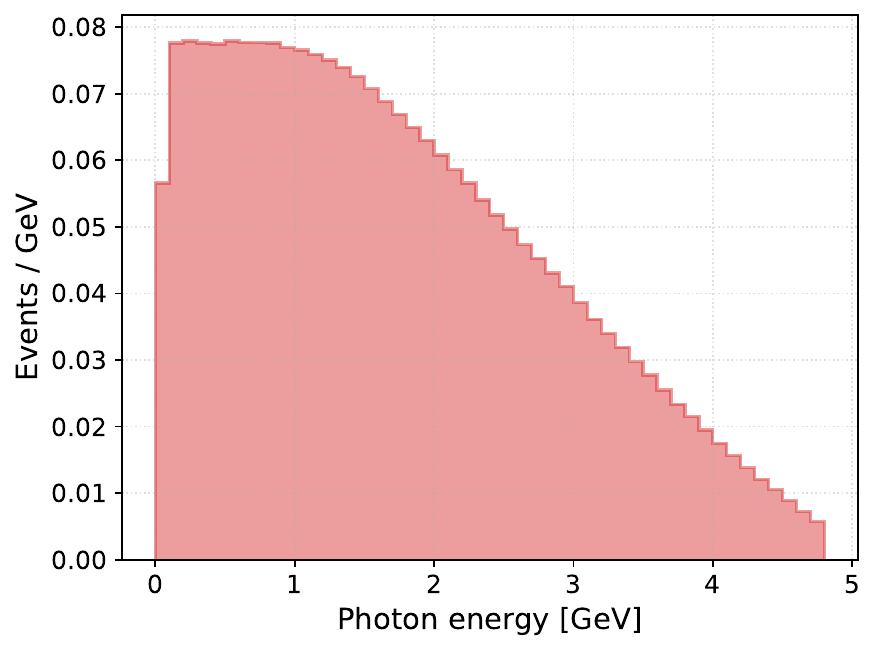}
  \caption{Photon energy distribution for composite heavy neutrino events for benchmark~A (left) and benchmark~B (right). The harder spectrum of benchmark~B reflects its larger heavy-neutrino mass.}
  \label{fig:em_shower_energy}
\end{figure}

The photon energy distribution, shown in Figure~\ref{fig:em_shower_energy}, is the most directly measured of the observables considered here. 
In the \(N\) rest frame the radiative decay is monochromatic, \(E_\gamma^\star = M_N/2\); boosting to the lab frame spreads this line roughly uniformly between zero and \(E_N\), so the photon carries on average half the heavy-neutrino energy. 
For benchmark~A the energy is degraded twice over. 
With \(\mu_{\rm IR}=100\)~MeV the entire flux is above production threshold, and the multi-body continuum at \(M_{\mathcal{U}}\geq 3\mu_{\rm IR}\) is readily populated, so a sizable fraction of the observed events come from fragmentation into three heavy neutrinos, of which the one decaying in the detector carries only about a third of the available energy. 
The resulting photon retains roughly a sixth of the incoming neutrino energy, and the spectrum peaks near \(E_\gamma\sim 0.15\)~GeV, well below the peak of the flux. 

Benchmark~B's spectrum is visibly harder, and for the opposite reason.
There the production step, rather than the decay, sets the energy scale. 
The heavier \(\mu_{\rm IR}=500\)~MeV closes the multi-body continuum for the Booster Neutrino Beam, so to a good approximation every event contains a single \(N\) inheriting essentially the full neutrino energy. 
More importantly, creating a 500~MeV state coherently requires the minimum momentum transfer, \(q_{\rm min}\simeq M_{\mathcal{U}}^2/2E_\nu\), which converts the compositeness scale into an effective flux threshold of \(E_\nu\gtrsim 2\)~GeV: only the energetic tail of the flux contributes at all. Together with the five-times-larger rest-frame photon energy, this yields a broad plateau between \(\sim 0.2\) and \(1\)~GeV, with a hard tail extending to several GeV.

\subsection{Angular Distribution of Photons}

\begin{figure}[tb]
  \centering
  \includegraphics[width=0.49\textwidth]{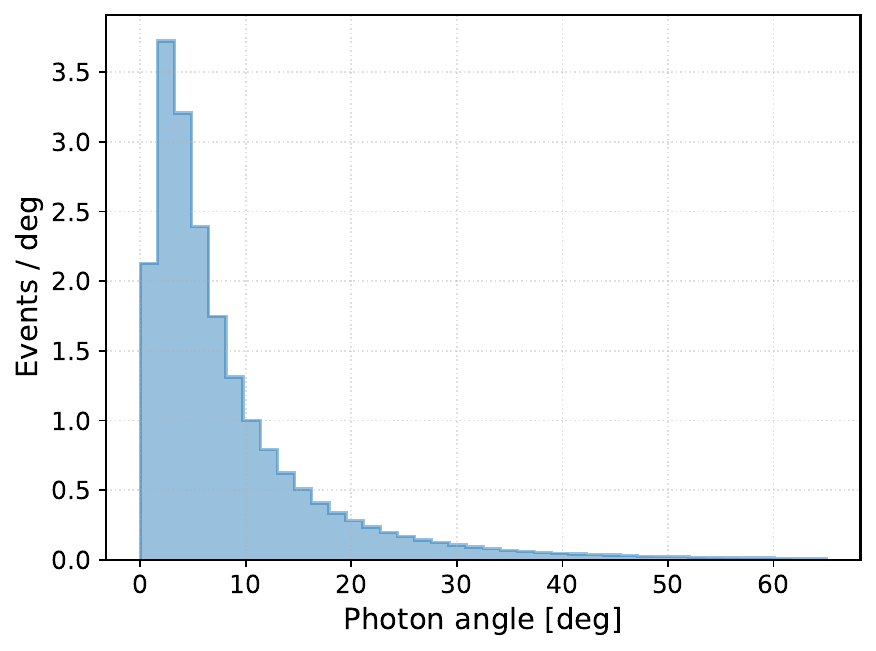}
  \includegraphics[width=0.49\textwidth]{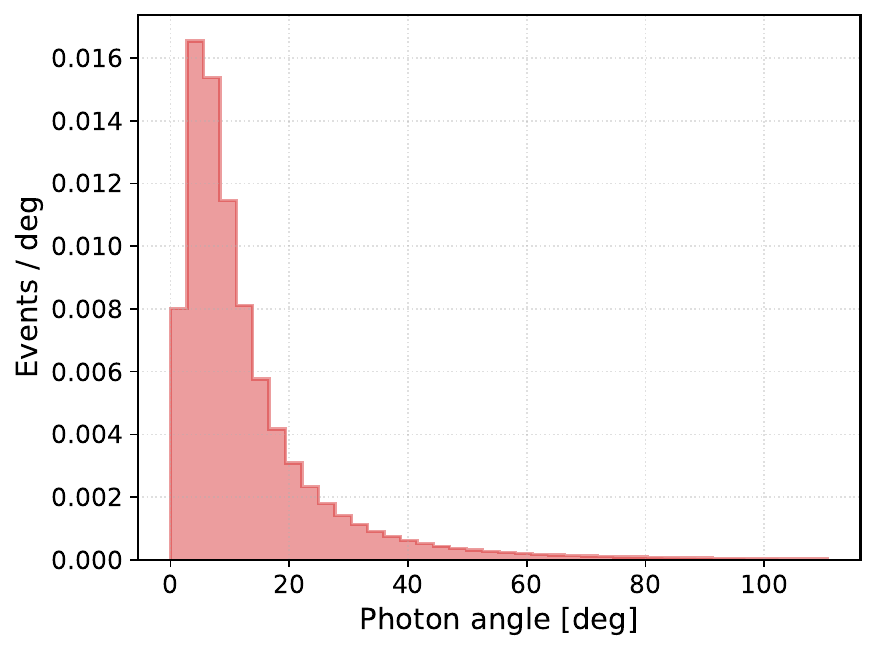}
  \caption{Angular distribution of the leading photon relative to the beam direction for benchmark~A (left) and benchmark~B (right), showing strong forward peaking.}
  \label{fig:cos_theta_leading}
\end{figure}

The angular distribution of the leading photon with respect to the beam direction, shown in Figure~\ref{fig:cos_theta_leading}, is controlled entirely by the decay boost. Coherent production keeps the momentum transfer below the inverse nuclear size, so the heavy neutrino emerges collinear with the beam, and folding the isotropic rest-frame decay with the boost of the parent yields a forward cone of characteristic opening \(\theta_\gamma \simeq M_N/E_N\). Since the decay shares the parent energy as \(E_N \simeq 2E_\gamma\) on average, this scale correlates directly with the spectra of Figure~\ref{fig:em_shower_energy}: for benchmark~A a \(100\)~MeV state against photon energies of a few hundred MeV peaks the distribution at \(\theta_\gamma \sim 3\)--\(5^\circ\), while for benchmark~B the five-times-larger mass is compensated by its correspondingly harder spectrum, displacing the peak only to \(\sim 5\)--\(8^\circ\), with a broader tail from the rare low-boost events. The collimation is generic rather than an accident of the benchmarks: light states are strongly boosted by any flux able to produce them, while heavy states are produced only above the coherence threshold.

\subsection{Discrimination from Neutrino--Electron Scattering}

\begin{figure}[tb]
  \centering
  \includegraphics[width=0.49\textwidth]{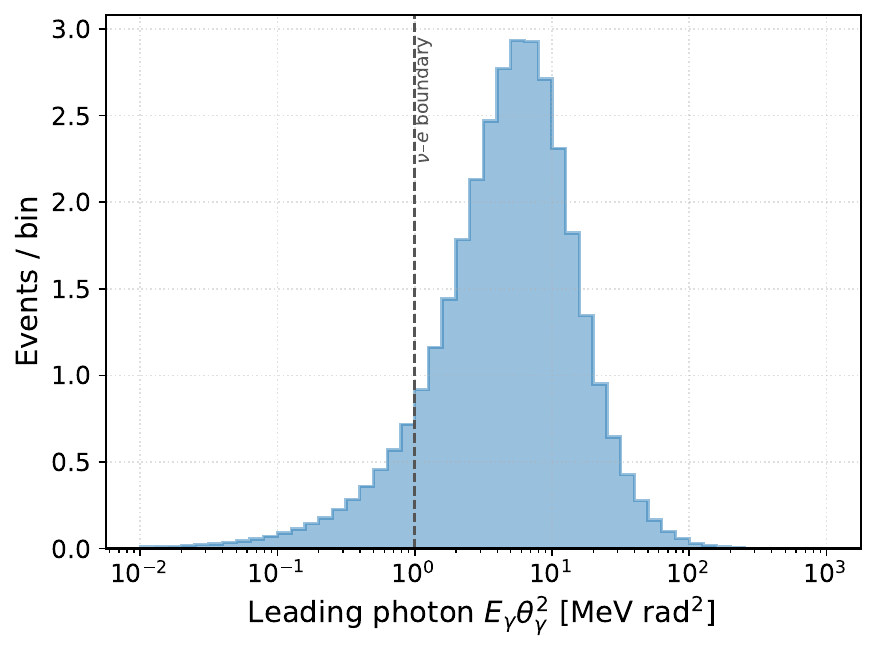}
  \includegraphics[width=0.49\textwidth]{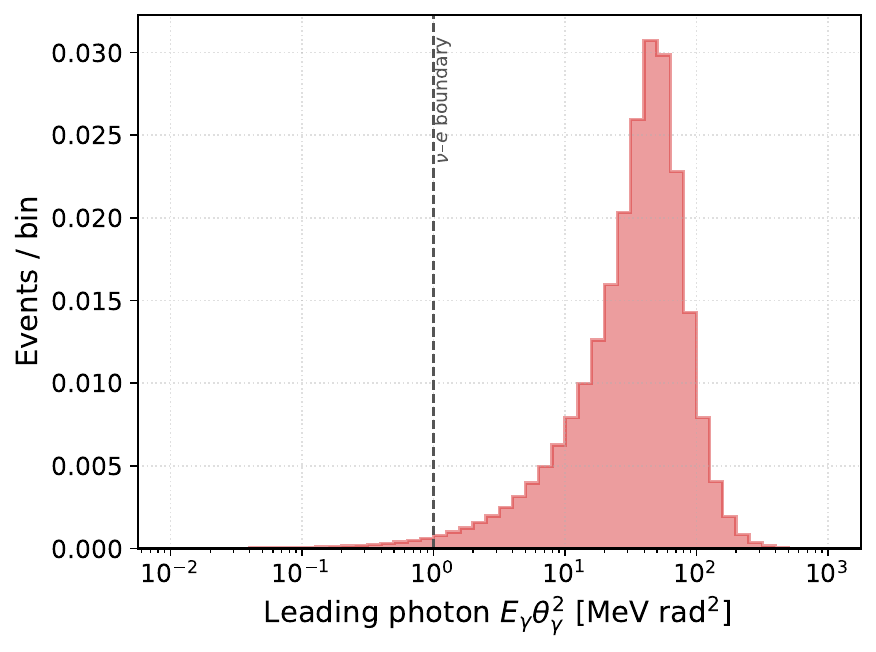}
  \caption{Distribution of \(E_\gamma\theta_\gamma^2\) for the leading photon for benchmark~A (left) and benchmark~B (right). The dashed line marks the kinematic boundary of neutrino--electron elastic scattering, \(E\theta^2 = 2m_e\). Note the logarithmic horizontal scale.}
  \label{fig:etheta2_leading}
\end{figure}

A single forward electromagnetic shower with no accompanying hadronic activity is also the topology of neutrino--electron elastic scattering, particularly in detectors that cannot separate photons from electrons. 
The standard handle on this process is the variable \(E\theta^2\): for elastic scattering on an electron at rest, kinematics confine the outgoing shower to \(E_e\theta_e^2 < 2m_e\). 
Our signal populates an entirely different region, as  shown in Figure~\ref{fig:etheta2_leading}. 
The photon distribution is localized around \(E_\gamma\theta_\gamma^2 \sim M_N^2/2E_N\), peaking 10--100 MeV\,rad$^2$, one to two orders of magnitude above the electron boundary. 
A simple \(E\theta^2\) selection therefore separates the signal from the neutrino--electron background while simultaneously providing yet another measure of the heavy-neutrino mass.

\section{Conclusion and Outlook}
\label{sec:conclusion}

We have investigated composite neutrino models where heavy neutrinos emerge as bound states from a near-conformal strongly coupled sector. The scaling dimensions of composite operators naturally suppress neutrino masses to sub-eV scales while maintaining compositeness scales \(\Lambda \sim 10~{\rm MeV}\)--\(1~{\rm GeV}\) accessible to experiments. Our central result is that the same composite operator responsible for the inverse-seesaw mass generates an enhanced transition dipole coupling \(d_{\mu N} \sim 10^{-6}\)--\(10^{-8}\,{\rm GeV}^{-1}\), parametrically above minimal Dirac or Majorana expectations. Through a dedicated event-level simulation of the production and radiative decay \(\nu X \to \mathcal{U} X \to \nu + n\gamma\, X\), we estimated experimental sensitivities directly within the model, taking MiniBooNE and MINERvA as concrete examples, and obtained predictions for the photon multiplicity, energy, and angular distributions, as well as for the \(E\theta^2\) discriminant against the neutrino--electron background.

For the benchmark scenarios studied here the observed radiative decays are predominantly single-photon, so the resulting event topology superficially resembles that of elementary models. 
The composite dynamics nevertheless leaves an imprint on the single-photon sample itself: fragmentation into several heavy neutrinos, of which only one decays inside the detector, softens the photon energy spectrum relative to a point-like heavy neutral lepton of the same mass. 
The genuinely multi-photon final states become accessible only once the multi-body unparticle continuum is populated, which requires lighter compositeness scales or harder production spectra than those probed here; in a realistic spectrum, heavier composite resonances can open even-multiplicity channels at still lower invariant mass~\cite{Ahmed:2025ldh}, further enhancing this signal.
Our benchmark parameters produce event characteristics broadly compatible with existing accelerator-neutrino data, though whether composite neutrinos bear on the reported short-baseline anomalies or face tension with astrophysical constraints requires further study.

These signatures are accessible to any neutrino experiment with single-photon reconstruction capabilities, and the multiplicity, energy, and angular distributions presented above provide concrete targets for such searches.
The natural next step is toward higher energies, where harder fluxes populate the multi-body continuum and reach heavier compositeness scales.
DUNE with its intense multi-GeV beam, FASER\(\nu\) and the LHC forward-physics program with TeV-scale collider neutrinos, and IceCube with atmospheric and astrophysical fluxes would progressively open the multi-photon regime as a qualitatively new probe of the composite dynamics.

{\bf Acknowledgments:} \ We are especially grateful to Zackaria Chacko for providing comments on the manuscript. We would also like to thank Abhish Dev, Zackaria Chacko, Paddy Fox and Matheus Hostert for helpful discussions and insightful comments. B.A. acknowledges support in part by the DOE grant DE-SC1019775, and the NSF grants OAC-2103889, OAC-2411215, and OAC-2417682. B.A.'s work was performed in part at the Aspen Center for Physics, with support by a grant from the Simons Foundation (1161654, Troyer).
P.M.'s work was supported by Fermi Forward Discovery Group, LLC under Contract No. 89243024CSC000002 with the U.S. Department of Energy, Office of Science, Office of High Energy Physics.

\bibliographystyle{JHEP}
\bibliography{refs}

\end{document}